\documentclass[lettersize,journal]{IEEEtran}
\usepackage{amsmath,amssymb,amsfonts}
\usepackage[T1]{fontenc}
\usepackage{algorithm} 
\usepackage{algorithmic}
\usepackage{mathrsfs}
\usepackage{array}
\usepackage{bm}
\usepackage{mathrsfs}
\usepackage[caption=false,font=normalsize,labelfont=sf,textfont=sf]{subfig}
\usepackage{textcomp}
\usepackage{stfloats}
\usepackage{url}
\usepackage{verbatim}
\usepackage{graphicx}
\usepackage{tikz}
\usepackage{booktabs}
\usepackage[colorlinks=true,citecolor=blue,urlcolor=blue]{hyperref}

\def\BibTeX{{\rm B\kern-.05em{\sc i\kern-.025em b}\kern-.08em
		T\kern-.1667em\lower.7ex\hbox{E}\kern-.125emX}}
\usepackage{cite}
\usepackage{amsthm}

\newtheorem{theorem}{Theorem}
\newtheorem{lemma}{Lemma}

\begin{document}

\title{A Lightweight Sensor Scheduler Based on AoI Function for Remote State Estimation over Lossy Wireless Channels\\
}

\author{
	Taige Chang, \IEEEmembership{Student Member, IEEE}, Xianghui Cao, \IEEEmembership{Member, IEEE} and Wei Xing Zheng, \IEEEmembership{Fellow, IEEE}
	%	 and Weibing Lu, \IEEEmembership{Senior Member, IEEE}
	\thanks{T. Chang and X. Cao are with the School of Automation, Southeast University, Nanjing 210096, China (E-mail: 220211789@seu.edu.cn; xhcao@seu.edu.cn). }
	\thanks{W. X. Zheng is with the School of Computer, Data and Mathematical Sciences, Western Sydney University, Sydney, NSW 2751, Australia (E-mail: w.zheng@westernsydney.edu.au). } 
	%\thanks{W. Lu is with the Center for Flexible RF Technology, State Key Lab of Millimeter waves, School of Information Science and Engineering, Southeast University, Nanjing 210096, China, and also with the MOE Frontiers Science for Mobile Information Communication and Security, Nanjing 210096, China (E-mail: wblu@seu.edu.cn). } 
}

\maketitle

\begin{abstract}
	This paper investigates the problem of sensor scheduling for remotely estimating the states of heterogeneous dynamical systems over resource-limited and lossy wireless channels. 
	Considering the low time complexity and high versatility requirements of schedulers deployed on the transport layer, we propose a lightweight scheduler based on an Age of Information (AoI) function built with the tight scalar upper bound of the remote estimation error. We show that the proposed scheduler is indexable and sub-optimal. We derive an upper and a lower bound of the proposed scheduler and give stability conditions for estimation error. Numerical simulations demonstrate that, compared to existing policies, the proposed scheduler achieves estimation performance very close to the optimal at a much lower computation time. 
\end{abstract}


\section{Introduction}

\IEEEPARstart{I}{n} many fields (e.g., health monitoring\cite{godkin2022feasibility}, power grid surveillance\cite{abdulwahid2021power}, UAV trajectory estimating\cite{mason2021remote} and so on), wirelessly and remotely estimating the states of dynamical systems are of great importance. 
Meanwhile, open-access wireless channels are often unreliable and bandwidth-limited, which in turn pose strict constraints on the performance of the remote estimation system\cite{akyildiz2002wireless}. And this further leads to a question of how to schedule wireless sensor transmissions to improve the estimation performance under those channel constraints\cite{wang2019whittle}. 

In the literature, periodic and event-triggered sensor schedulers are the most prevalent ones\cite{shi2011time}. 
Periodic schedulers plan the transmission time instants offline. It was shown that the optimal scheduling policy for the remote sensor scheduling problem of a scalar estimation system that schedules the sensors for fixed scheduling time instants across a finite time horizon is to make a uniform transmission decision of the measurement \cite{yang2011deterministic}.
For a more complicated two-system scheduling problem in which only one is allowed to transmit at each decision iteration, the optimal solution is a similar periodic policy that alternatively schedules the two systems for a given number of time instants\cite{Shi_2012}. 
Ren \emph{et al.}\cite{Zhu_Ren_2013} studied the sensor scheduling problem of a general linear time-invariant system under energy constraints. They further proposed a periodic scheduling policy and derived a sufficient stability condition. However, without taking real-time estimation performance into account, offline policies cannot perform on-demand transmissions. Also, the stability requirement of the remote estimation system often leads the periodic schedulers to transmit more than event-triggered ones\cite{shi2014event}, which consumes more network resources. 

Event-triggered policies achieve a trade-off between the estimation performance and the transmission load by designing online schedulers based on the real-time status of the estimation systems. 
The pioneering work of Astrom \emph{et al.} \cite{astrom2002comparison} showed that in contrast to the periodic scheduling policies (viewed as a Riemann sampling of a stochastic process), there should exist a kind of scheduling policies derived from the Lebesgue sampling, which are nowadays called the event-triggered policies. Since then, researchers have proposed many event-based scheduling policies, among which one of the most intuitive ones is an event-triggering condition based on the gap between the prediction and the measurement\cite{shi2014event}. 
Instead of designating certain phenomena as the ``event'' for triggering transmissions, a more natural way is to derive the event by solving a convex optimization problem\cite{leong2016sensor}. In \cite{yang2015deterministic}, a sub-optimal scheduling policy was obtained by solving the convex optimization problems derived from a convex upper bound of the expected remote estimation error. 
Han \emph{et al.}\cite{han2015stochastic} argued that, unlike the periodic schedules, the randomness induced by the network destroys the Gaussianity of the estimated states in the event-triggered cases, which renders the estimation problem essentially intractable. To this end, they proposed an event-triggered stochastic scheduler whose decision variable follows a Gaussian distribution. 
Although the event-triggered stochastic policies based on a Gaussian-distributed random variable solve the above Gaussianity issue, under the same transmission environment, there always exist deterministic policies that perform better than the stochastic ones\cite{yu2021stochastic}. 

When the sensors directly transmit their measurements, the remote estimation error covariance may evolve nonlinearly over time and be difficult to analyze\cite{sinopoli2004kalman}. Also, the random information losses caused by the network lead to non-negligible performance degradation of the Kalman filter running remotely compared with the one deployed locally. 
To this end, Hovareshti \emph{et al.}\cite{hovareshti2007sensor} proposed the smart sensors that run the Kalman filter locally and transmit the estimated state to the remote estimator, which allows the latter to update the estimated state linearly. They also proved that under this system setting, the remote estimator is the optimal one. 
By transforming the maximum likelihood estimation problem of the remote estimation into a quadratic optimization problem, Shi \emph{et al.}\cite{shi2014event} showed that the optimal remote Kalman filter equipped with smart sensors and event-triggered schedulers will form a  time-varying Riccati equation. 
The above deterministic event-triggered policies only use the information of the current time step, which may cause a loss of information during scheduling. This issue further led to a  stochastic-deterministic hybrid event-triggered policy\cite{Yu_2022}. 
Nevertheless, the timely system status required by the event-triggered policies dramatically increases the transmission and computational burden of the scheduler. 

Recently, the notion of Age of Information (AoI) emerges as a metric that measures the obsoleteness of the information received from the source\cite{kaul2012real}. 
Researchers from the communication community have proposed many AoI-based schedulers to improve data freshness, such as the Whittle-index-based policy\cite{kadota2018scheduling}, the deep reinforcement learning-based policy\cite{leong2020deep}, the threshold greedy policy\cite{arafa2021timely}, and the truncated-threshold policy\cite{Tang_2020}.
Other schedulers based on UoI (Urgency of Information derived from the AoI) were also proposed\cite{zheng2020urgency, chen2022uncertainty}.  
However, regarding sensor scheduling tasks in remote estimation systems, directly optimizing the AoI is ignorant of the system dynamics and hence is undesirable. For example, the performance of the policy that greedily schedules transmissions to minimize the AoI is dramatically worse than that of minimizing an AoI function derived from the remote estimation error\cite{ayan2019age}.  
Kl{\"u}gel \emph{et al.} \cite{klugel2019aoi} tackled a single sensor scheduling problem and proved that the optimal scheduling policy based on the AoI holds a threshold structure.  
Wang \emph{et al.}\cite{wang2019whittle} addressed a multi-sensor scheduling problem with limited channel constraints and proposed an AoI function-based Whittle index policy by formulating the estimation error as an AoI function. The above scheduling policies based on the AoI function require much matrix calculations with high computation complexity, which is undesirable for a scheduler deployed on the transport layer.  

In this paper, we propose a lightweight scheduling policy deployed on the transport layer of the network to minimize the multi-sensor remote estimation error under channel restrictions. Based on the Whittle index policy, the proposed lightweight scheduling policy makes decisions according to the Whittle indexes constructed by a scalar AoI function deduced from the characteristic parameters of the dynamical systems. We show that the proposed policy has superior computation efficiency and versatility in implementation. Analyzing a constraint-relaxed optimization problem, we derive a closed-form expression of the Whittle indexes and a lower bound of the proposed policy. 
%Further, we prove the indexability of the lightweight scheduling policy that promises its sub-optimality. 
By reconstructing the proposed scheduling policy using the Lyapunov energy function, we obtain an upper bound for system performance when our policy is applied. The stability conditions are also established. 
The contributions of our work are summarized as follows: 
\begin{enumerate} 
	\item We propose a lightweight sensor scheduler based on the tight scalar upper bound of the remote estimation error.
	\item We prove the indexability of the proposed lightweight scheduling policy that promises sub-optimality. 
	\item We derive a necessary and a sufficient stability conditions for the estimation error stability. 
	\item We analyze the performance of the lightweight scheduling policy by giving its upper and lower bounds. 
\end{enumerate} 

The remainder of this paper is laid out as follows. Section \ref{section-system-model} presents the system model and formulates the optimal scheduling problem. Section \ref{section-scheduling-policies} proposes the lightweight scheduling policy. Section \ref{section-performance-analysis} derives an upper and a lower bound of our policy as well as the necessary and sufficient stability conditions. Section \ref{section-simulation} presents the simulation results, followed by conclusions in Section \ref{conclusion}. 

\emph{Notation:} $ \mathbb{C}_{N}^{M}$ denotes the number of combinations of selecting $ M $ objects from total $ N $ objects. $ \mathbb{P}(\cdot) $ and $ \mathbb{P}_{X}(\cdot) $ denote the probabilities of a stochastic event and a stochastic variable $ X $, respectively. $ \Lambda_A $ and $ U_A $ denote the Jordan canonical form and the corresponding transformation matrix of $ A $, respectively. $ \mathbb{E}[\cdot] $ stands for the expectation of a random variable while $ \mathbb{E}[\cdot|\cdot] $ represents the conditional expectation. The superscript $ T $ denotes the matrix/vector transpose operation. $ \mathrm{Tr}(\cdot) $ stands for the trace of a square matrix, while $ \|\cdot\| $ represents the Euclidean norm. $ \rho(\cdot) $ and $\lambda_{\min}(\cdot)$ denote the spectral radius and the minimum eigenvalue of a square matrix, respectively. $diag\{\cdots\}$ denotes a diagonal matrix. 
$ w.p. $ is the abbreviation for ``with probability''. $ \mathbb{N}^+ $ and $ \mathbb{N} $ denote the sets of all positive and non-negative integers, respectively. $ \mathbb{R} $ and $ \mathbb{R}^{+} $ denote the sets of real numbers and positive real numbers, respectively. $ \emptyset $ stands for the empty set.

\section{Problem Formulation\label{section-system-model}}

\subsection{System Model}

\begin{figure} 
	\centering
	\includegraphics[width=0.48\textwidth]{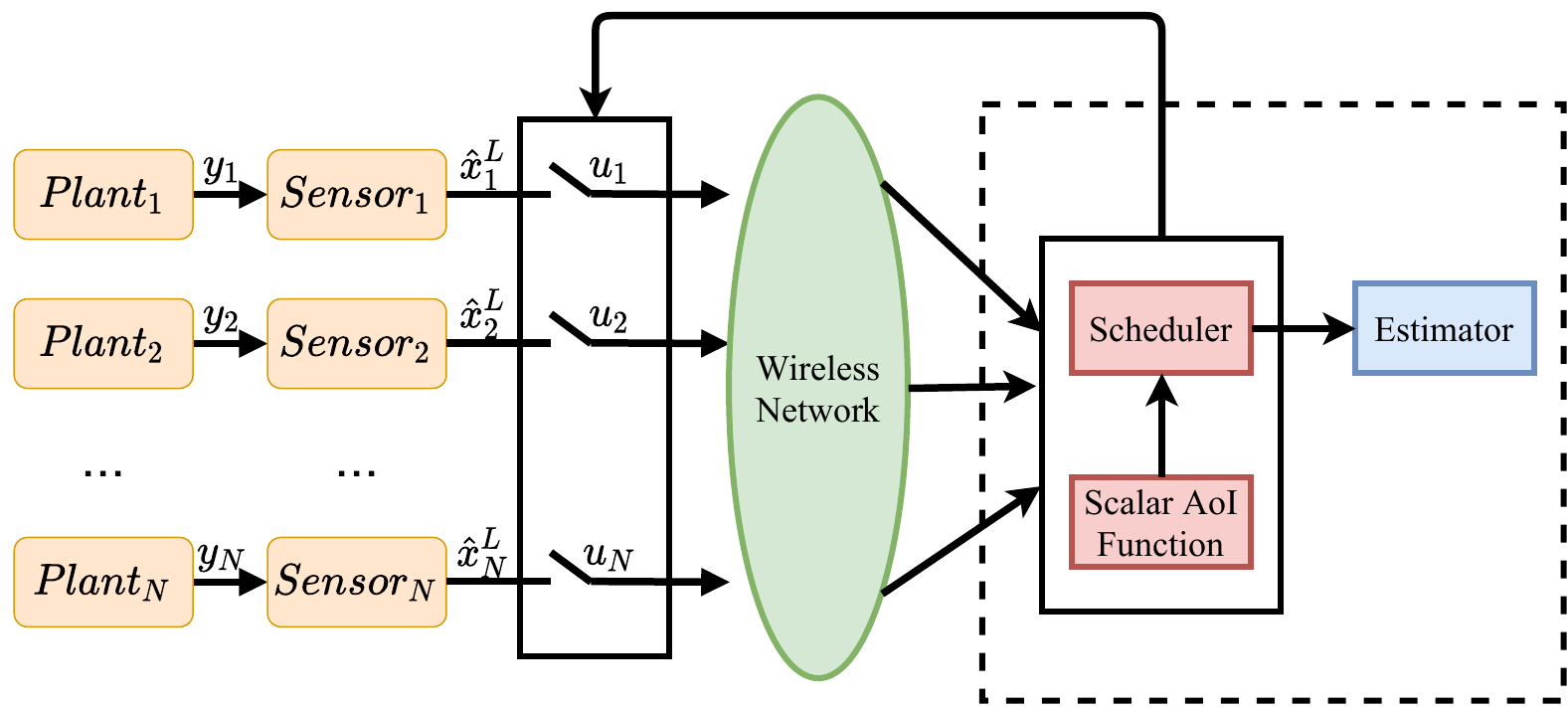}
	\caption{The remote estimation system model. \label{fig-system-model}}
\end{figure}

As shown in Fig. \ref{fig-system-model}, we consider an estimator remotely estimating $ N $ plants, indexed by $ i\in \mathcal{N}\triangleq \{1, 2, \cdots, N\} $. Each plant $ i $ is modeled as a discrete-time linear time-invariant system as follows: 
\begin{align}
	x_i(t+1)=A_ix_i(t)+\omega_i(t), \label{kalman-filter-state} 
\end{align}
where $t$ is the discrete-time step, $ x_i\in \mathbb{R}^{n_i} $ is the plant state, and $ \omega_i $ is the plant noise. A corresponding smart sensor is installed for each plant $ i $ to measure the state $x_i$. The measurement $ y_i \in\mathbb{R}^{m_i}$ is modeled as:
\begin{align}
	y_i(t)=C_ix_i(t)+v_i(t), \label{kalman-filter-observe} 
\end{align}
where $v_i$ denotes the measurement noise, $w_i$ and $v_i$ are assumed to be independent Gaussian noises with zero means and covariance matrices $Q_i$ and $R_i$, respectively. 
We also assume that each pair of $ (A_i, C_i) $ is observable, $ (A_i, \sqrt{Q_i}) $ is controllable, and $ \rho(A_i) > 1 $\cite{you2011mean}. At time step $t\in \mathbb{N}$, after taking the measurement, each smart sensor obtains a local estimate of the corresponding plant's state and then sends the estimate to the remote estimator through some shared wireless communication channels. 

Denote by $ \mathbb{I}_i(t) \triangleq \{y_i(1), y_i(2), \cdots, y_i(t)\} $ the information  sensor $ i $ can access till time $ t $, and further define \emph{a priori} and \emph{a posterior} local estimates respectively as follows: 
\begin{align}
	\hat{x}^L_i(t|t-1) &\triangleq{} \mathbb{E}[x_i(t)|\mathbb{I}_i(t-1)] , \nonumber\\
	\hat{x}^L_i(t|t) &\triangleq{} \mathbb{E}[x_i(t)|\mathbb{I}_i(t)].\nonumber
\end{align}
Denote $P^L_i(t|t-1)$ and $P^L_i(t|t)$ as the corresponding error covariances of $\hat{x}^L_i(t|t-1)$ and $\hat{x}^L_i(t|t)$, respectively.
Then, the optimal estimate of each system $ i $ in terms of mean square sense, as known to all, is obtained by the standard Kalman filter as follows\cite{anderson2012optimal}: 
\begin{subequations}
	\begin{align}
		\hat{x}^L_i(t|t-1)={}&A_i\hat{x}^L_i(t-1|t-1),\label{kalman filter}\\
		P^L_i(t|t-1)={}&A_iP^L_i(t-1|t-1)A_i^T+Q_i,\label{kalman-filter-p predict}\\
		\hat{x}^L_i(t|t)={}&\hat{x}^L_i(t|t-1)+K_i(t)(y_i(t)\nonumber\\ &-C_i\hat{x}^L_i(t|t-1)),\label{kalman-filter-x-predict}\\
		K_i(t)={}&P^L_i(t|t-1)C_i^T(C_iP^L_i(t|t-1)C_i^T+R_i)^{-1},\label{kalman-filter-p-update}\\
		P^L_i(t|t)={}&P^L_i(t|t-1)-K_i(t)C_iP^L_i(t|t-1). \label{kalman-filter-k}
	\end{align}\label{kalman-filter}% 
\end{subequations}
The initial states of each local Kalman filter are specified as follows: $\forall i\in \mathcal{N}, \hat{x}^L_i(0|0) $ is a Gaussian random variable with known mean $ \bar{x}_i(0|0) $ and known covariance $ P_i(0|0) $ \cite{anderson2012optimal}.
Given that the Kalman filter converges exponentially fast, we assume that the error covariance has already converged to its steady-state value $\{\bar{P}_i, i\in \mathcal{N}\}$ in the following\cite{liu2021remote}. 
Note that $ \{Q_i\} $ and $ \{\bar{P}_i\} $ are all positive definite matrices. 

A scheduler is employed to control the channel access of the sensors. We assume that due to limited communication resources, only a limited number of sensors, say $ M $, can get channel access and transmit data to the remote estimator at each time step $ t\in \mathbb{N} $, i.e., 
\begin{align}
	\sum_{i=1}^N {u_i(t)\le M}, \forall t\in \mathbb{N}, 
\end{align}
where $ u_i(t)\in \{0, 1\} $ denotes the scheduling decision of sensor $ i $. The scheduler schedules sensor $i$ to send its local estimate $ \hat{x}^L_i(t|t) $ to the estimator at time step $t$ if $ u_i(t) = 1 $, and otherwise if $u_i(t) = 0$. 
We define the decision vector at time step $t$ as $ \boldsymbol{u}(t)\triangleq [u_1(t), \cdots, u_N(t)] $, and the decision space as $ \mathcal{U} \triangleq \{0, 1\}^N $. 
We assume that each sensor can complete the transmission within a single time step. 
In consideration of the typically unreliable nature of wireless channels, we denote the successful transmission indicator of each sensor $ i $ as $s_i(t)\in \{0, 1\} $, and the successful transmission probability as $ \mathbb{P}(s_i(t) = 1) = p_i $. 
To simplify the analysis, we make the assumption that the first transmission after system initialization is successful\cite{Yu_2022}.

Based on whether the local estimate $\hat{x}^L_{i}(t|t)$ is successfully transmitted at time step $t$, the remote estimator runs a filter to obtain the estimate of $x_i(t)$, as described in detail in \cite{hovareshti2007sensor}. 
We assume that each successful transmission from a sensor can be completed within a single time step. Because the remote estimator updates the remote state at the end of each time step, the remotely estimated state is at least one time step behind the local one. 
Therefore, we follow \cite{liu2021remote} and perform a one-step prediction upon successful transmission, as the following equations show: 
\begin{align}
	\hat{x}_i(t)=\left\{
	\begin{array}{ll}
		A_i\hat{x}^L_{i}(t-1|t-1), &\text{if } \gamma_i(t)=1,\\
		A_i\hat{x}_i(t-1), &\text{otherwise}, 
	\end{array}
	\right.
\end{align}
where 
\begin{align}
	\gamma_i(t) = s_i(t) u_i(t). \label{gamma-definition}
\end{align}
Thus, the estimation error covariance of the remote estimator becomes
\begin{align}
	P_i(t)={}\left\{
	\begin{array}{ll}
		A_i\bar{P}_iA_i^T, &\text{if }\gamma_i(t)=1, \\
		A_iP_i(t-1)A_i^T+Q_i, &\text{otherwise. }
	\end{array}
	\right. \label{smart-P}
\end{align}

\subsection{Problem Formulation}
The main objective of this paper is to design a lightweight scheduling policy (in terms of $ \{\boldsymbol{u}(t)\} $) to minimize the long-term estimation error at the estimator. 
%In many existing studies, the scheduler makes decisions based on the real-time estimation error and is deployed locally, which is computationally costly. We propose a novel lightweight scheduling policy that works on the network layer in this paper. 

The AoI of sensor $i$'s data at time step $t$, denoted by $\Delta_i(t)$, is defined as the number of time steps elapsed from the latest time when the local estimate was successfully transmitted. Hence, the AoI evolves as\cite{kadota2018scheduling}
\begin{align}
	\Delta_i(t)=\left\{
	\begin{array}{ll}
		1, &\text{if } \gamma_i(t)= 1, \\
		\Delta_i(t-1)+1, &\text{otherwise}.
	\end{array}
	\right.\label{AoI}
\end{align}
Note that when $ \gamma=1 $, we set $ \Delta=1 $ to take into account the time of the successful transmission in terms of the number of time steps. By combining (\ref{smart-P}) and (\ref{AoI}), we obtain the following expression for the remote estimation error covariance: 
\begin{align}
	P_i(t)={}A_i^{\Delta_i(t)}\bar{P}_i(A_i^T)^{\Delta_i(t)} + \sum_{k = 0}^{\Delta_i(t) - 1}A_i^kQ_i(A_i^T)^k. \label{P-i}
\end{align}
Then we can formulate the original optimization problem of the remote estimation system as follows: 
\begin{subequations}
	\begin{align}
		\min_{\left\{\boldsymbol{u}(t)\in \mathcal{U}\right\}}\quad &J_{origin}\triangleq \lim_{\tau\rightarrow \infty}\frac{1}{\tau}\mathbb{E}
		\left[ \sum_{t=1}^\tau\sum_{i=1}^N \textrm{Tr}(P_i(t))\right], \label{trivial-optimal-problem-a}\\
		\text{s.t. }\quad &\sum_{i=1}^N {u_i(t)\le M}. \quad \forall t\in \mathbb{N}. \label{trivial-optimal-problem-b}
	\end{align}\label{trivial-optimal_problem}%
\end{subequations}

Solving the online scheduling problem above typically requires evaluating the objective function in real-time, which can result in high computational complexity. Alternatively, many existing schedulers that determine $ \{\boldsymbol{u}(t)\} $ based on AoI are ignorant of the plant dynamics and undoubtedly achieve poor performance in solving the above problem (\ref{trivial-optimal_problem}). To address these issues, we propose a lightweight scheduling policy that makes decisions based on a scalar AoI function constructed with system characteristic parameters. Specifically, we define the AoI function of system $ i $ as follows:
\begin{align}
	f_i(\Delta_i(t)) \triangleq \beta_i\alpha_i^{\Delta_i(t)}, \label{f} 
\end{align}
where $ \alpha_i $ and $ \beta_i $ are the characteristic parameters of system $ i $ that satisfy the following inequalities: 
\begin{align}
	\textrm{Tr}(A_i^k\bar{P}_i(A_i^k)^T)&\le{} f_i( k),\label{inequality-fp} \\
	\textrm{Tr}(A_i^kQ_i(A_i^k)^T)&\le{}f_i(k).\label{inequality-fq} 
\end{align} 

The meaning of the parameters $\alpha_i$ and $\beta_i$ can be interpreted as follows: $\alpha_i$ and $\beta_i$ indicate the rate and scale, respectively, that the estimation error regarding system $i$ increases when no data packages from sensor $i$ are available to the remote estimator. 
%Note that $ \{f_i(\Delta_i(t)), i\in \mathcal{N}\} $ leads to an upper bound of the remote estimation error, which can be easily verified by substituting (\ref{inequality-fp}) and (\ref{inequality-fq}) into (\ref{P-i}) as follows: 
\begin{align}
	\textrm{Tr}(P_i(t))\le \sum_{k = 1}^{\Delta_i(t)} f_i(k). \nonumber
	%	\label{inequality-p-f} 
\end{align} 

There are many choices of $ \{\alpha_i\} $ and $ \{\beta_i\} $ that satisfy inequalities (\ref{inequality-fp}) and (\ref{inequality-fq}). However, as the following Lemma shows, under the choices of $ \{\alpha_i\} $ and $ \{\beta_i\} $, the upper bounds (\ref{inequality-fp}) and (\ref{inequality-fq}) are the tightest ones. 
\begin{lemma}
	\label{lemma-trace-inequalities}
	For any positive definite matrices $ Q_i $ and $ \bar{P}_i $ and $ \forall k\in \mathbb{N} $, the following inequalities hold:  
	\begin{subequations}
		\begin{align}
			\textrm{Tr}(A_i^kQ_i(A_i^T)^k)&\le {} \rho^{2k}(A_i)\textrm{Tr}(Q_i), \tag{13a}\label{trace-ineq-Q}\\ 
			\textrm{Tr}(A_i^k\bar{P}_i(A_i^T)^k)&\le {} \rho^{2k}(A_i)\textrm{Tr}(\bar{P}_i).  \tag{13b}\label{trace-ineq-P}
		\end{align}\label{trace-ineq}% 
	\end{subequations} 
	Furthermore, if letting $ \alpha_i =  \alpha_i^* = \rho^2(A_i) $ and $ \beta_i = \beta_i^* = \max\{\textrm{Tr}(A_i\bar{P}_iA_i^T)/\alpha_i^*, \textrm{Tr}(Q_i)\} $, then there is a tight upper bound of the remote estimation error covariance $P_i(t)$, i.e., $\forall i \in \mathcal{N}, \forall t > 0, \textrm{Tr}(P_i(t)) \le \beta_i^* \frac{(\alpha_i^*)^{\Delta_i(t)}-1}{\alpha_i^*-1}$. 
\end{lemma} 
	\begin{IEEEproof} 
		Since for any square matrices $ D $ and $ E $ of the same dimension, $ \textrm{Tr}(DE) = \textrm{Tr}(ED) $,
		we get that
		\begin{align}
			\textrm{Tr}\left(A_i^kQ_i(A_i^T)^k\right) \nonumber
			= & \textrm{Tr}\left(A_i^TA_iA_i^{k-1}Q_i(A_i^T)^{k-1}\right). 
		\end{align} 
		Considering that $ (A_i^T)^kA_i^k $ and $A_i^{k-1}Q_i(A_i^T)^{k-1}$ are all symmetric and positive-definite matrices, we conclude that \cite{fang1994inequalities}: 
		\begin{align}
			&\textrm{Tr}(A_i^kQ_i(A_i^k)^T) \nonumber\\ 
			\le{} &\rho^{2}(A_i)\textrm{Tr}(A_i^{k-1}Q_i(A_i^T)^{k-1}) \nonumber\\
			\le{} &\rho^{2}(A_i)\rho^{2}(A_i)\textrm{Tr}(A_i^{k-2}Q_i(A_i^T)^{k-2}) \nonumber\\
			\le{} &\cdots 
			\le{} \rho^{2k}(A_i)\textrm{Tr}(Q_i).  \nonumber
		\end{align} 	
		In a similar argument, we can prove the left inequality of (\ref{trace-ineq-Q}) as well as (\ref{trace-ineq-P}).  
		
		In the following, we prove by contradiction that the choice of $ \alpha_i =  \alpha_i^* = \rho^2(A_i) $ and $ \beta_i = \beta_i^* = \max\{\textrm{Tr}(A_i\bar{P}_iA_i^T)/\alpha_i^*, \textrm{Tr}(Q_i)\} $ yields a tight upper bound of the remote estimation error covariance. Suppose that there exists $\alpha_i^\prime < \alpha_i^*$ such that 
		\begin{align}
			\textrm{Tr}(A_i^kX_i(A_i^k)^T) \le \beta_i(\alpha_i^\prime)^k < \beta_i (\alpha_i^*)^k, \forall k > 0, \label{inequality-alpha-prime}
		\end{align} 
		where $X_i$ stands for either $\bar{P}_i$ or $Q_i$. Apparently, $X_i>0$. 
		Denote the Jordan canonical form associated with the matrix $A_i$ as $J_i$, where $A_i = U_iJ_iU_i^{-1}$ and $U_i$ is an invertible matrix. Considering that the largest absolute value of the diagonal elements of the matrix $\frac{1}{\rho(A_i)}J_i$ is $1$, we have $\lim_{k\to\infty} (\frac{1}{\alpha^{*}})^\frac{k}{2}A^k_i =\lim_{k\to\infty}U(\frac{1}{\alpha_i^{*}})^{k/2}J_i^kU^{-1}=\tilde{A}_i \neq 0$. On the other hand, 
		it can be easily verified from (14) that  
		\begin{align}
			&\textrm{Tr}(\tilde{A}_iX_i\tilde{A}_i^T) \nonumber \\={}&  \lim_{k\rightarrow \infty}\textrm{Tr}\left((\frac{1}{\sqrt{\alpha_i^*}}A_i)^kX_i(\frac{1}{\sqrt{\alpha_i^*}}(A_i)^T)^k\right) \nonumber \\={} &\left(\frac{1}{\alpha_i^{*}}\right)^k 
			\lim_{k\rightarrow \infty}\textrm{Tr}(A_i^kX_i(A_i^k)^T)
			\nonumber\\  \le{}& 
			\lim_{k\rightarrow \infty}\beta_i^*(\frac{\alpha_i^\prime}{\alpha_i^*})^k = 0, \label{inequality-alpha-prime-rho}
		\end{align} 
		where the last equality holds due to the fact that $\alpha_i^\prime < \alpha_i^*$. 
		Since $X_i$ is positive definite, we obtain that $ \textrm{Tr}(\tilde{A}_iX_i\tilde{A}_i^T)> 0 $, which contradicts with the inequality in (\ref{inequality-alpha-prime-rho}). Hence, the tightness of $\alpha_i^*$ is proved. Furthermore, it is clear that when $t = 0$, $\beta_i^* = \max\{\textrm{Tr}(A_i\bar{P}_iA_i^T)/\alpha_i^*, \textrm{Tr}(Q_i)\}$ is tight, and for $t > 0$, $\beta_i^*$ and $\alpha_i^*$ still lead to an upper bound of $P_i(t)$. Thus, we complete the proof. 
	\end{IEEEproof} 	

In the rest of this paper, we will focus on the following optimization problem, which aims to minimizing the long-term average of the AoI function: 
\begin{subequations}
	\begin{align}
		\min_{\left\{\boldsymbol{u}(t)\in \mathcal{U}\right\}}\quad &J\triangleq \lim_{\tau\rightarrow \infty} J(\tau), \label{optimal-problem-a}\\
		\text{s.t. }\quad &\sum_{i=1}^N {u_i(t)\le M}, \quad \forall t\in \mathbb{N},  \label{optimal-problem-b}
	\end{align}\label{optimal_problem}%
\end{subequations}
where $ J(\tau) $ is defined as follows: 
\begin{align}
	J(\tau) \triangleq \frac{1}{\tau}\mathbb{E}
	\left[ \sum_{t=1}^\tau\sum_{i=1}^N f_i(\Delta_i(t))\right]. 
\end{align}

\section{Lightweight Scheduling Policy\label{section-scheduling-policies}}

In this section, we introduce our lightweight scheduling policy. 
We propose a relaxed version of the problem (\ref{optimal_problem}) and the corresponding decoupled problem that shares the optimal solution. 
We further prove that the optimal solution to the relaxed problem has a threshold structure, which induces a Whittle index formula that plays a central role in our policy. 
%\miold{
	%\subsection{The Lightweight Policy}}
We observe that (\ref{optimal-problem-a}) is formally inconsistent with (\ref{optimal-problem-b}) since the former takes the expectation over a long period while the latter is restrictive at each single time step, making the problem (\ref{optimal_problem}) hard to solve. Therefore, we consider the following relaxed problem: 
\begin{subequations}
	\begin{align}
		\min_{\{\boldsymbol{u}(t)\in \mathcal{U}\}}\quad &J, \label{relaxed_optimal_problem objective}\\
		\text{s.t. }\quad &\lim_{\tau\rightarrow \infty} \frac{1}{\tau}\mathbb{E}\left[ \sum_{t=1}^\tau\sum_{i=1}^N u_i(t)\right] \le M. \label{relaxed_optimal_problem a}
	\end{align}\label{relaxed_optimal_problem}%
\end{subequations}
Denote the objective function $ J $ of the problems (\ref{optimal_problem}) and (\ref{relaxed_optimal_problem}) under their optimal policies by $ J_{opt} $ and $ J_{relaxed} $, respectively. Since the problem (\ref{relaxed_optimal_problem}) relaxes the constraints of the problem (\ref{optimal_problem}), the following inequality holds: 
\begin{align}
	J_{relaxed} \le J_{opt}. 
	%	\label{inequality-J}
\end{align} 

	\begin{lemma}
		The optimal policy and the corresponding optimal objective function of the problem (\ref{relaxed_optimal_problem}) are identical to those of the following decoupled optimization problem, respectively: 
		\begin{subequations}
			\begin{align}
				\min_{\{\boldsymbol{u}(t)\in \mathcal{U}\}}\quad & J, \\
				\text{s.t. }\quad &
				\lim_{\tau\rightarrow \infty} \frac{1}{\tau}\mathbb{E}\left[\sum_{t=1}^\tau u_i(t)\right] \le m^u_i\label{decoupled a}, \quad \forall i\in \mathcal{N}, 
			\end{align}\label{decoupled_relaxed_optimal_problem}%
		\end{subequations}
		where $ m_i^u $ denotes the average scheduled transmission time instants of sensor $ i $ as follows: 
		\begin{align}
			m_i^u = \lim_{\tau\rightarrow \infty} \frac1\tau \mathbb{E}\left[\sum_{j=1}^\tau u_i^{relaxed}(j)\right], \label{definition-m-i-u}
		\end{align} 
		where $ \{u_i^{relaxed}(t)\} $ denotes the optimal policy to the problem (\ref{relaxed_optimal_problem}). 
	\end{lemma} 
	
	\begin{IEEEproof} 
		First, we observe that exchanging the decisions of the problem (\ref{relaxed_optimal_problem}) over time (i.e., letting $ u_i(t) = u_i(t^{'}) $ and $ u_i(t^{'}) = u_i(t) $, for any $ t\neq t^{'} $) will not violate the constraint (\ref{relaxed_optimal_problem a}), while the objective function (\ref{relaxed_optimal_problem objective}) is changed. However, in the following we confirm by contradiction that the optimal objective functions of the problems (\ref{relaxed_optimal_problem}) and (\ref{decoupled_relaxed_optimal_problem}) are equal to each other. We denote the optimal objective functions of problems (\ref{relaxed_optimal_problem}) and (\ref{decoupled_relaxed_optimal_problem}) as $J$ and $J^\prime$ and the corresponding optimal policies as $\mathbf{u}_i$ and $\mathbf{u}^\prime_i$, respectively. 
		
		Suppose that $J > J^\prime$. Since $\mathbf{u}^\prime_i$ is optimal, it is easy to verify that 
		\begin{align} 
			\lim_{\tau\rightarrow \infty} \mathbb{E}\left[\frac{1}{\tau}\sum_{\tau} u^\prime_{i}(\tau)\right] = m_i^u, 	\forall i\in \mathcal{N}.   
		\end{align}
		Otherwise by increasing the left-hand-side (LHS) of (\ref{decoupled a}), $J^\prime$ could be further reduced, which contradicts to the optimality of $J^\prime $. Furthermore, considering that $\{m_i^u, i\in \mathcal{N}\}$ is chosen by (\ref{definition-m-i-u}), the following equality holds:
		\begin{align}
			\lim_{\tau\rightarrow \infty} \mathbb{E}\left[\frac{1}{\tau}\sum_{\tau} u_{i}(\tau)\right] = 
			\lim_{\tau\rightarrow \infty} \mathbb{E}\left[\frac{1}{\tau}\sum_{\tau} u^\prime_{i}(\tau)\right]. 
		\end{align} 
		Also, considering that the objective functions of the problems (\ref{relaxed_optimal_problem}) and (\ref{decoupled_relaxed_optimal_problem}) are of the same form, we can always modify $\mathbf{u}_i$ and $\mathbf{u}^\prime_i$ to be the same and reduce $J$, which would lead to a contradiction to the optimality of $J$. We could prove similarly for the case $J < J^\prime$ and conclude that $J = J^\prime$ for sure. Thus, we complete the proof.
	\end{IEEEproof}
Next, we prove that the optimal solution to the problem (\ref{decoupled_relaxed_optimal_problem}) holds a threshold structure. 
\begin{theorem}[The threshold structure] 
	\label{theorem decouple}
	The optimal policies of the problem (\ref{decoupled_relaxed_optimal_problem}) hold a threshold structure, i.e., $ \forall i \in \mathcal{N} $, 
	\begin{align}
		u^{*}_i(t) = \left\{
		\begin{array}{ll}
			1, &\text{if } \Delta_i(t) < \Delta_{i, \text{th}}, \\
			0, &\text{otherwise}, 
		\end{array}
		\right.\label{threshold-scheduler}
	\end{align}
	where $ \Delta_{i, \text{th}} $ is the threshold associated with system $ i $. Moreover, the same threshold structure holds for the optimal solution to the problem (\ref{relaxed_optimal_problem}). 
\end{theorem} 
\begin{IEEEproof}
	%	The proof is given in Appendix \ref{proof theorem decouple}. 
		We first introduce a set of Lagrangian multipliers $ \{\mathcal{W}_i, i\in \mathcal{N}\} $ associated with the constrains of the decoupled problem (\ref{decoupled a}) and obtain the following Bellman equation of problem (\ref{decoupled a}): 
		\begin{align}
			V_i^d(\Delta_i)+\theta_i^d 
			= \min_{u_i\in \{0, 1\}}\Big\{&
			(1-p_iu_i)V_i^d(\Delta_i+1)\nonumber\\&+f_i(\Delta_i)+\mathcal{W}_iu_i\Big\}. 
			\label{threshold-scheduler-simplified-Bellman-equation}
		\end{align}
		
		We first assume that the optimal policy holds a threshold structure, i.e., equation (\ref{threshold-scheduler}) holds. 
		Given that $ \Delta_i < \Delta_{i, \text{th}} $ for any $ i $ under the above assumption, the optimal decision is $ u_i=0 $. Therefore, we can conclude that the expected gain of decision $ u_i=1 $ in the Bellman equation (\ref{threshold-scheduler-simplified-Bellman-equation}) is greater than that of $ u_i = 0 $. This can be shown by the following inequality:
		\begin{subequations}
			\begin{align}
				\mathcal{W}_i+(1-p_i)V_i^d(\Delta_i)&>{}V_i^d(\Delta_i),  \label{threshold-scheduler-inequality-value-function-1} \\
				\mathcal{W}_i+(1-p_i)V_i^d(\Delta_i+1)&<{}V_i^d(\Delta_i+1),  \label{threshold-scheduler-inequality-value-function-2}
			\end{align} %
		\end{subequations}
		where equation (\ref{threshold-scheduler-inequality-value-function-2}) is derived for the case $ \Delta_i\ge \Delta_{i, \text{th}} $. 
		Then, we recursively obtain the following expression of $ V_i^d(\Delta_i) $: 
		\begin{align}
			&V_i^d(\Delta_i)\nonumber\\=&\left\{
			\begin{array}{ll}
				\frac{\beta_i(\alpha_i^{\Delta_{i, \text{th}}} - \alpha_i^{\Delta_i})}{\alpha_i-1} + \Delta_i\theta_i^d + V_{i, \text{res}}, 
				&\text{if } \Delta_i < \Delta_{i, \text{th}},  \\ 
				\frac{\beta_i\alpha_i^{\Delta_i}}{1-\alpha_i+p_i\alpha_i} + \frac{\mathcal{W}_i-\theta_i^d}{p_i}, &\text{otherwise}, 
			\end{array}
			\right.\label{threshold-scheduler-V}
		\end{align}
		where $ V_{i, \text{res}} \triangleq V_i^d(\Delta_{i, \text{th}})-\Delta_{i, \text{th}}\theta_i^d  $.  
		
		Finally we confirm that the threshold structure assumption is consistent with (\ref{threshold-scheduler-V}) by writing the difference equation of $ V_i^d(\Delta_i) $.
		Please refer to Appendix \ref{proof theorem decouple} for the details of the proof. 
\end{IEEEproof}

Now we turn back to the problem (\ref{optimal_problem}). Since the optimal threshold scheduling policy may violate constraint (\ref{optimal-problem-b}), we introduce the \emph{Whittle index} derived from the above optimal thresholds to guarantee that constraint. 
Hereby, the Whittle index represents the urgency of a decision, which is defined as the Lagrangian multiplier such that both decisions $ u_i=1 $ and $ u_i=0 $ yield the same expected objective function value. In other words, as the Whittle index increases, it is more urgent to make the decision $ u_i = 1 $. Thus, our policy schedules the sensors with the largest $ M $ Whittle indexes to obtain a sub-optimal $ J $.

Before applying the Whittle index, the \emph{indexability}, which promises the asymptotic optimality, should be confirmed. That is, when the Whittle index increases, if the passive set (the set of AoIs that lead to decision $ u_i = 0 $) expands monotonically from $ \emptyset $ to $ \mathbb{N}^+ $, then the policy is indexable\cite{maatouk2020optimality}. 

\begin{theorem}[Whittle index and indexability]
	\label{theorem whittle index}
	The Whittle index of any sensor $ i\in \mathcal{N} $ regarding the problem (\ref{decoupled_relaxed_optimal_problem}) is given by 
	\begin{align}
		%W_i(\Delta_i) = {}&\frac{\beta_ip_i^2\Delta_{i}\alpha_i^{\Delta_{i} + 1}}{1+\alpha_ip_i - \alpha_i} - \frac{\beta_ip_i\alpha_i(\alpha_i^{\Delta_{i}} - 1)}{\alpha_i - 1}.   \label{W} \\
		W_i(\Delta_i) ={} &\beta_ip_i\alpha_i^{\Delta_i+1}\left(\frac{p_i\Delta_{i}}{1+\alpha_ip_i - \alpha_i}-\frac{1}{\alpha_i - 1}\right) \nonumber\\&+\frac{\beta_ip_i\alpha_i}{\alpha_i - 1}. \label{W}
	\end{align}
	Moreover, the indexability holds under (\ref{W}). 
\end{theorem}
	\begin{IEEEproof}
		%	The proof is given in Appendix \ref{appendix whittle index}. 
		Remember that the Whittle index is defined as the critical value of the Lagrangian multiplier such that both decisions $ u_i = 0 $ and $ u_i = 1 $ obtain the same expected value. 
		By substituting (\ref{threshold-scheduler-V}) into (\ref{threshold-scheduler-inequality-value-function-1}) and making some simplification, we get: 
		\begin{align}
			\mathcal{W}_i \le{} &\frac{\beta_ip_i^2\Delta_{i, \text{th}}\alpha_i^{\Delta_{i, \text{th}} + 1}}{1+\alpha_ip_i - \alpha_i} - \frac{\beta_ip_i\alpha_i(\alpha_i^{\Delta_{i, \text{th}}} - 1)}{\alpha_i - 1}. \label{theorem whittle index W}
		\end{align}
		Replacing $ \Delta_{i, \text{th}} $ with $ \Delta_i $ and simplifying the right-hand side (RHS) of (\ref{theorem whittle index W}), we obtain the Whittle index in (\ref{W}). To see the proof in detail, please refer to Appendix \ref{appendix whittle index}. 
	\end{IEEEproof}

%\minew{Based on Theorem \ref{theorem whittle index}, we proposed a lightweight scheduling policy that schedules the systems that hold the largest $M$ Whittle's index.}
Based on Theorem \ref{theorem whittle index}, we summarize the proposed lightweight scheduling policy as Algorithm \ref{algorithm-scalar-whittle-index}. 
\begin{figure}[htbp]
	\renewcommand{\algorithmicrequire}{\textbf{Input:}}
	\renewcommand{\algorithmicensure}{\textbf{Output:}}
	\begin{algorithm}[H]
		\caption{The lightweight scheduling policy\label{algorithm-scalar-whittle-index}}
		\begin{algorithmic}[1]
			\REQUIRE $\{\alpha_i\}, \{\beta_i\}, \{p_i\}, M, N, \{\Delta_{i}(t)\}$. 
			\ENSURE The lightweight scheduling result $ \{u_i\} $ at time $ t $. 
			\FOR {$ i = 0 $ to $ N $}        
			\STATE Calculate $ W_i $ by equation (\ref{W}); 
			\ENDFOR
			\STATE Sort $ \{W_i, i\in \mathcal{N}\} $ in descending order; 
			\STATE Choose $ \{i_{1}, \cdots, i_{M}\} $ such that $ \{W_{i_1}, \cdots, W_{i_M}\} $ are the largest $ M $ Whittle indexes;  
			\STATE Return the scheduling results $ \{u_i=1|i\in \{i_{1}, \cdots, i_{M}\}\} \cup \{u_i=0|i\notin \{i_{1}, \cdots, i_{M}\}\}$.
		\end{algorithmic}
	\end{algorithm}
\end{figure}
%\subsection{Discussions\label{policies-discussions}}
Since the most time-costing operation of the proposed scheduler is the sorting of the Whittle indexes, we conclude that the time complexity of the proposed policy is $ O(N\log(N)) $. 
Considering that the estimation error-based policies, e.g., \cite{ayan2019age, wang2019whittle}, require matrix calculations which incur additional time complexity of $ O(Nn_{\max}^3)$ ($ n_{\max} \triangleq \max_{i\in \mathcal{N}} n_i $), the proposed policy dramatically reduces the computation time.

The advantages of our lightweight scheduling policy can be summarized as easy deployment, low computational complexity, high versatility, and secure service provisioning. 
\begin{enumerate}
	\item Our lightweight scheduler works on the transport layer of the remote estimator to schedule the sensors, which conforms to the hierarchical structure of the network and makes it easy to deploy.
	\item Since only a few parameters and simple scalar calculations are required, the computational complexity of our lightweight scheduling policy is very low, which allows for implementation with a large system scale and a strict limit on computation time.  
	\item Our lightweight scheduling policy has better versatility. Considering that when the plants change, we merely need to adjust the system characteristic parameters ($ \alpha_i $ and $ \beta_i $). On the contrary, the traditional scheduling policies require the entire plants' models and real-time estimation error information, which is more plant-specific. 
	\item Since our lightweight scheduling policy does not require calculating or storing the real-time estimation information, the scheduler does not need to decrypt the sensor data, if encrypted, which reduces the risk of information at the scheduler itself. 
\end{enumerate}

\section{Performance Analysis\label{section-performance-analysis}}

In this section, we analyze the performance of our lightweight scheduling policy in solving the problem (\ref{optimal_problem}). Denote by $ J_{lightweight} $ the objective function $ J $ of the problem (\ref{optimal_problem}) achieved by the proposed policy.  
We characterize the performance using upper and lower bounds of $ J_{lightweight} $. These bounds have explicit expressions based on the structural parameters such as $ \{p_i, i\in \mathcal{N}\} $, $ \{\alpha_i, i\in \mathcal{N}\} $, and $ \{\beta_i, i\in \mathcal{N}\} $. Note that $ J_{opt} $ also lies in between those bounds, thus, they indicate the gap between our policy and the optimal one, i.e., 
\begin{align} 
	\underline{J} \le J_{opt} \le J_{lightweight} \le \bar{J},   \nonumber
\end{align} 
where $ \underline{J} $ and $ \bar{J} $ are the lower and upper bounds, respectively.
%\subsection{Lower Bound}

We first look at the lower bound.
\begin{theorem}[Lower bound]
	\label{theorem-lower-bound}
	If $ \alpha_i(1-p_i) < 1 $, then the lower bound $ \underline{J} $ is given by
	\begin{align}
		\underline{J} ={} &\sum_{i = 1}^N\mathcal{K}_i\frac{p_i\alpha_i^{\Delta_{i, \text{th}}^{*}}-\alpha_ip_i+\alpha_i-1}{\Delta^{*}_{i,\text{th}}p_i+1-p_i}, \label{lower-bound}
	\end{align} 
	where $ \mathcal{K}_i=\frac{p_i\alpha_i\beta_i}{(\alpha_i-1)(1-\alpha_i+\alpha_ip_i)} $, and $ \{\Delta^{*}_{i, \text{th}}\} $ are the thresholds that minimize $ \underline{J} $. 
	
	Similarly, the lower bound of $ J_{origin} $, denoted by $ \underline{J}_{origin} $, is given by 
	\begin{align} 
		\underline{J}_{origin} ={} &\sum_{i = 1}^N\hat{\mathcal{K}}_{i}\frac{p_i\rho^{2\hat{\Delta}^{*}_{i, \text{th}}}(A_i)-\rho^2(A_i)p_i+\rho^2(A_i)-1}{\hat{\Delta}^{*}_{i, \text{th}}p_i+1-p_i}, \label{lower-bound-origin}
	\end{align} 
	where $ \{\hat{\Delta}^{*}_{i, \text{th}}\} $ denote the thresholds that minimize $ \underline{J}_{origin} $ and 
	\begin{align} 
		\hat{\mathcal{K}}_{i} &= \frac{p_i\rho^2(A_i)\zeta_i\min\left\{\lambda_{\min}(Q_i), \lambda_{\min}(\bar{P}_i)\right\}}{(\rho^2(A_i)-1)(1-\rho^2(A_i)+\rho^2(A_i)p_i)}, \nonumber \\
		\zeta_i &= \lambda_{\min}(A_i)(U_{A_i}(U_{A_i}^T))\lambda_{\min}(A_i)(U_{A_i}^{-1}({U_{A_i}^{-1}})^T). \nonumber
	\end{align} 
\end{theorem} 
\begin{IEEEproof}
	By regarding the AoI of each system $ i $ as the state $ \{\Delta_i(t), t\in \mathbb{N}\} $, we reformulate the problem (\ref{relaxed_optimal_problem}) into $ N $ discrete Markov decision processes (DMDP) and analyze them similarly. Based on Theorem \ref{theorem decouple}, we depict the state transition diagram of the DMDP in Fig. \ref{fig-dmdp}. 
	
	\begin{figure}
		\centering
		\includegraphics[width=0.48\textwidth]{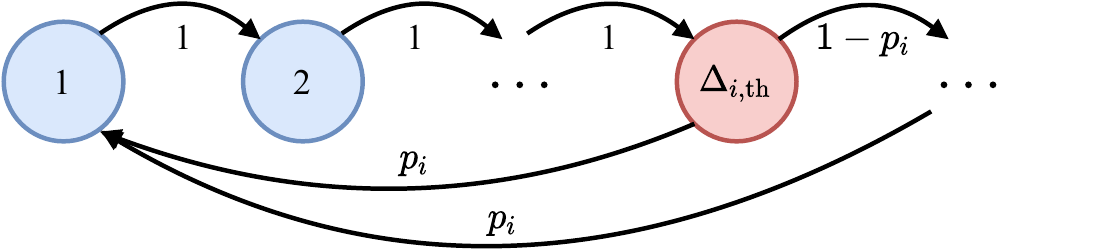}
		\caption{The state transition diagram of the Markov chain under DMDP. \label{fig-dmdp}}
	\end{figure}
	As we have assumed that the first transmission is successful, the AoI $\Delta_i(t)$ is independent and identically distributed (i.i.d). According to \cite{xie2009stability}, it follows the following distribution:
	\begin{align}
		\label{dmdp-stable-distribution}
		\Psi_i(\Delta_i(t))=\left\{
		\begin{array}{ll}
			\frac{p_i}{\Delta_{i, \text{th}}p_i+1-p_i}, \qquad \text{if } \Delta_i(t) < \Delta_{i, \text{th}},\\
			\frac{p_i(1-p_i)^{\Delta_i(t)-\Delta_{i, \text{th}}}}{\Delta_{i, \text{th}}p_i+1-p_i}, 
			\text{otherwise}. 
		\end{array} \right. 
	\end{align}
	Then the problem (\ref{relaxed_optimal_problem}) can be reformulated
	%	\miold{\footnote{The details are given in Appendix \ref{appendix-lower-bound}. }} 
	as follows:
	\begin{subequations} 
		\begin{align}
			\min_{\{\Delta_{i, \text{th}}\in \mathbb{N}^+\}}\quad &\sum_{i = 1}^N\sum_{\Delta_i=1}^\infty f_i(\Delta_i(t))\Psi_i(\Delta_i(t)), \label{dmdp-optimal-problem-a}\\
			\text{s.t. }\quad &\sum_{i = 1}^N\frac{1}{\Delta_{i, \text{th}}p_i+1-p_i}\le M. \label{dmdp-optimal-problem-b}
		\end{align}\label{dmdp-optimal-problem}%
	\end{subequations} 
	Consequently, the lower bounds (\ref{lower-bound}) and (\ref{lower-bound-origin}) can be proved by applying the inequalities (\ref{trace-ineq}), which are detailed in Appendix \ref{appendix-lower-bound}. 
\end{IEEEproof} 

The problem (\ref{dmdp-optimal-problem}) is a discrete optimization problem that is generally difficult to find the exact solution. However, 
there exists an upper bound for each optimal threshold $ \Delta_{i, \text{th}}^{*} $, as the next lemma shows. 

%\minew{In the next Lemma, we show that there exits an upper bound for each optimal threshold $ \Delta_{i, \text{th}}^{*} $. }

\begin{lemma}
	\label{lemma-searching-set} 
	The following inequality holds: 
	\begin{align}
		\Delta_{i, \text{th}}^{*}&\le \frac{1}{p_i}\left(\frac{1}{\mathcal{G}_{i, \min}} + 2p_i - 1\right), \quad \forall i\in\mathcal{N},  \label{lemma-upperbound} 
	\end{align}
	where $ \forall i\in\mathcal{N} $, $ \mathcal{G}_{i, \min} $ is defined as
	\begin{subequations}
		\begin{align}
			\mathcal{G}_{i, \min} ={} &M - \max_{\{\Delta_{j, \text{th}}\in \mathbb{N}^{+}\}}  \sum_{j\in \mathcal{N}, j\neq i} \frac{1}{\Delta_{j, \text{th}}p_j+1-p_j}, \nonumber\\ 
			&\text{s.t. }\qquad  \sum_{j\in \mathcal{N}, j\neq i} \frac{1}{\Delta_{j, \text{th}}p_j+1-p_j} < M. \nonumber
		\end{align}  % 
	\end{subequations} 
\end{lemma}
\begin{IEEEproof}
	The proof is given in Appendix  \ref{appendix-searching-set}. 
\end{IEEEproof}
%\subsection{Upper Bound}

As far as we know, the upper bound of the policies based on the Whittle index is challenging. Jiang \cite{jiang2021analyzing} analyzed the Partial Derivative Equation (PDE) of the AoI by assuming that scheduling decisions are Poisson distributed. However, it is rare for deterministic policies to follow this distribution. On the other hand, rebuilding the Whittle index using a Lyapunov energy function can be an effective method for finding an upper bound of the performance of the policy based on the Whittle index \cite{kadota2018scheduling}. To this end, we adopt a Lyapunov energy function of the following form:
\begin{align} 
	\mathcal{L}(\boldsymbol{\Delta}(t)) ={}& \sum_{i=1}^N \mathcal{L}_i(\Delta_i(t))\nonumber\\ \triangleq{}& \sum_{i=1}^Nl_{i1}\Delta_i(t)\beta_i\alpha_i^{\Delta_i(t)} + l_{i2}\beta_i\alpha_i^{\Delta_i(t)}, \label{L} 
\end{align} 
where $ \{\mathcal{L}_i(\Delta_i(t))\} $ denotes the Lyapunov energy function of system $ i $, $ l_{i1} $ and $ l_{i2} $ are designable coefficients, and $ \boldsymbol{\Delta}(t) \triangleq [\Delta_1(t), \cdots, \Delta_N(t)]$. The corresponding Lyapunov drift $ \vec{\mathcal{L}}(t) $ is defined as follows: 
\begin{align}
	\vec{\mathcal{L}}(t) \triangleq{}& \mathbb{E}\big[\mathcal{L}(\boldsymbol{\Delta}(t+1)) - \mathcal{L}(\boldsymbol{\Delta}(t)) | \boldsymbol{\Delta}(t)\big]. \label{definition-lyapunov-drift}
\end{align} 

To derive an upper bound of the Lyapunov drift function (\ref{definition-lyapunov-drift}) in the following Lemma. We first reformulate (\ref{definition-lyapunov-drift}) as follows\footnote{The derivation of this formula is detailed in Appendix \ref{appendix-upper-bound}}: 
\begin{align}
	\vec{\mathcal{L}}(t) = &  \sum_{i = 1}^N -\mathbb{E}[\gamma_i(t)|\Delta_i(t)]\Big\{l_{i1}\alpha_i\Delta_i(t)\beta_i\alpha_i^{\Delta_i(t)} \nonumber\\& - \alpha_i(l_{i1} + l_{i2}) + \alpha_i(l_{i1} + l_{i2})
	\beta_i\alpha_i^{\Delta_i(t)}\Big\} 
	\nonumber\\& +\Big\{l_{i1}(\alpha_i-1) \Delta_{i}(t)\beta_i\alpha_i^{\Delta_i(t)} \nonumber\\& + (l_{i1}\alpha_i+l_{i2}\alpha_i - l_{i2})\beta_i\alpha_i^{\Delta_i(t)} \Big\},  \label{theorem-upper-bound-lyapunov-drift}
\end{align}
where $ \mathbb{E}[\gamma_i(t)|\Delta_i(t)] $ represents the conditional successful transmission rate under a scheduling policy based on AoI.

According to Kim \emph{et al.}\cite{kim2014scheduling}, the Lyapunov drift (\ref{theorem-upper-bound-lyapunov-drift}) is minimized by greedily scheduling $ M $ sensors that have the largest $ G_i $s, i.e., the first term in the summation of (\ref{theorem-upper-bound-lyapunov-drift}) with $ \mathbb{E}[\gamma_i(t)|\Delta_i(t)] $ replaced with $ p_i $: 
\begin{align}
	G_i = &{}p_i\beta_i\Big[l_{i1}\alpha_i\Delta_i(t)\alpha_i^{\Delta_i(t)} + (l_{i1} + l_{i2}) 
	\alpha_i(\alpha_i^{\Delta_i(t)} - 1)\Big].  \nonumber
\end{align}

Now, we are ready to derive the upper bound of the Lyapunov drift (\ref{definition-lyapunov-drift}), which is one of the fundamental parts in the derivation of the upper bound of $ J_{lightweight} $. 
\begin{lemma}
	\label{lemma-lyapunov-drift-inequality} 
	The following inequality holds: 
	\begin{align} 
		\vec{\mathcal{L}}(t) \le{} & \sum_{i = 1}^N -q^{*}_iG_i 
		+\beta_i\alpha_i^{\Delta_i(t)}\Big[l_{i1}(\alpha_i - 1) \Delta_{i}(t) \nonumber\\&+ l_{i1}\alpha_i+l_{i2}\alpha_i - 	l_{i2}\Big], \label{inequality-lyapunov-drift} 
	\end{align}	
	where $ \{q_i^{*}, i\in \mathcal{N}\} $ is the optimal solution to the following problem: 
	\begin{subequations}
		\begin{align}
			\min_{\{\boldsymbol{q}\in \mathcal{Q}\}}\quad &\sum_{i = 1}^N\frac{\beta_i(\alpha_i-1)}{1-\alpha_i+\alpha_ip_iq_i},\label{problem-statical-scheduler-a}\\
			s.t.\quad 
			&\sum_{i=1}^N q_i \le M, \label{problem-statical-scheduler-b}\\
			&\alpha_i (1 - p_iq_i) < 1, \quad \forall i\in \mathcal{N} , \label{problem-statical-scheduler-c} \\
			&q_i\in (0, 1), \quad \forall i\in \mathcal{N}.   \label{problem-statical-scheduler-d} 
		\end{align}\label{problem-statical-scheduler}%
	\end{subequations} 
	In the above, $ \boldsymbol{q}\triangleq [q_1, \cdots, q_N] $ denotes the decision vector and $ \mathcal{Q} \triangleq (0, 1]^N $ represents the decision space.
\end{lemma}
\begin{IEEEproof}
	The proof is given in Appendix \ref{appendix-lyapunov-drift-inequality}. 
\end{IEEEproof}

By carefully designing the parameters $l_{i1}, l_{i2} $, we can further obtain an upper bound of $ J_{lightweight} $. 

\begin{theorem}[Upper bound]
	\label{theorem-upper-bound}
	If the following condition holds: 
	\begin{align} 
		\sum_{i = 1}^N\frac{1}{p_i}(1-\frac{1}{\alpha_i}) &< M, \label{upper-bound-existing-condition} 
	\end{align} % 
	then an upper bound of $ J_{lightweight} $ under the proposed policy is given by  
	\begin{align} 
		&\bar{J} = \frac{\mathcal{C} + \sum_{i = 1}^N p_iq^{*}_i\beta_i\alpha_i(l_{i1} + l_{i2})}{\min_{i\in \mathcal{N}}\{\eta_i \tilde{\Delta}^*_i-\mathcal{S}_i\}},  \label{upper-bound}
	\end{align} 
	where $ \forall i\in \mathcal{N} $, 
	\begin{subequations}
		\begin{align} 
			l_{i1} &\triangleq \frac{p_i}{1-(1-p_i)\alpha_i},\label{l1}\\
			l_{i2} &\triangleq \frac{\alpha_i-1+p_i-2\alpha_ip_i}{(\alpha_i-1)[1-\alpha_i(1-p_i)]}, \label{l2} \\
			\tilde{\Delta}^*_i &={} \mathop{\arg\min}_{\eta_i \Delta-S_i>0} \left\{\eta_i \Delta-S_i\right\}, \label{theorem-upperbound-h} \\ 
			\eta_i &={} l_{i1}[1-\alpha_i(1-p_iq_i^{*})],  \\
			\mathcal{S}_i &={} \alpha_i(l_{i1}+l_{i2})(1-p_iq_i^{*})-l_{i2}, \\ 
			\mathcal{C} &={} 
			\left\{\begin{array}{ll} 
				\sum_{i=1}^N \eta_i \tilde{\Delta}^*_i\beta_i\alpha_i^{\tilde{\Delta}_i^*}, &\text{if } \tilde{\Delta}_i^* > 1, \\
				0, &\text{otherwise}. 
			\end{array} \right. \label{theorem-upperbound-c}
		\end{align} % 
	\end{subequations}
\end{theorem} 

\begin{IEEEproof}
	%	The proof is given in Appendix \ref{appendix-upper-bound}. 
		By selecting $ l_{i1}$ and $ l_{i2}$ as specified in (\ref{l1}) and (\ref{l2}), respectively, for all sensors $ i\in \mathcal{N} $, we can confirm that $ G_i $ matches the Whittle index in (\ref{W}). In this scenario, the policy that minimizes the Lyapunov drift (\ref{theorem-upper-bound-lyapunov-drift}) yields the same performance as our lightweight scheduling policy. Thus, by analyzing the Lyapunov drift (\ref{theorem-upper-bound-lyapunov-drift}), we can obtain an upper bound of $ J $. For the details of the proof, please refer to Appendix \ref{appendix-upper-bound}.
\end{IEEEproof}
%\miold{ 
	%Note that by choosing $ \{l_{i1}, i\in \mathcal{N}\} $ and $ \{l_{i2}, i\in \mathcal{N}\} $ as (\ref{l1}) and (\ref{l2}), respectively, (\ref{L}) represents the Lyapunov energy of the remote estimation system since the optimization policy that minimizes (\ref{lyapunov-drift}) performs the same as our lightweight one. 
	%
	%\subsection{Stability Conditions\label{stability-condition}}}

Note that if inequality (\ref{upper-bound-existing-condition}) does not hold, then there does not exist a random scheduling policy that could stabilize the remote estimation system. 

The stability of the remote estimation error is affected by two factors: the scheduling policy and the successful transmission rate. In order to ensure stable estimation, a certain level of sensor data arrivals at the remote estimator is required. This implies that the following stability condition must be met.
\begin{theorem}[Necessary stability condition] 
	\label{theorem-necessary-condition-of-system-stability}
	If the remote estimation system is stable in the mean square sense, i.e., $ J_{origin} < \infty $, then the following inequality holds: 
	\begin{align} 
		\rho^2(A_i)(1-p_i)<1, \quad \forall i\in \mathcal{N}. \label{stability-condition-1} 
	\end{align} 
\end{theorem} 
\begin{IEEEproof}
	%	Given that $ J_{origin} < \infty $, we can conclude that each system $ i $ is stable. Additionally, we observe that the transmission decision $ \{u_i(t)=1, \forall t\in \mathbb{N}\} $ is the most stable for system $i$ under any scheduling policy (including our lightweight one). In other words, if system $i$ is not stable under the decision $ \{u_i(t)=1, \forall t\in \mathbb{N}\} $, it will not be stable under any other policies either. Based on this observation, we derive the necessary stability condition for system $i$ as given in (\ref{stability-condition-1}) \cite[Theorem 3]{you2011mean}, completing the proof. 
	Please refer to Appendix \ref{appendix-theorem-necessary-condition-of-system-stability} for the details of the proof. 
\end{IEEEproof}

In the following theorem, we derive a sufficient condition for the stability of the remote estimation system. 

\begin{theorem}[Sufficient stability condition] 
	\label{theorem-sufficient-condition-of-system-stability}
	A sufficient stability condition for the remote estimation system is as follows: 
	\begin{align} 
		\rho^2(A_i)(1-q_i^{*}p_i)<1, \quad \forall i\in \mathcal{N}, \label{stability-condition-2} 
	\end{align} 
	where $ \{q_i^{*}, i\in \mathcal{N}\} $ is the optimal solution to the problem (\ref{problem-statical-scheduler}). 
\end{theorem} 

\begin{IEEEproof}
	%	Since the randomized policy described above renders an upper bound of the estimation error, a sufficient condition for estimation stability under the optimal randomized policy is also sufficient for stability under the proposed policy. Under the randomized policy, we have: 	
	%	\begin{align}
		%		\mathbb{P}(u_i(t))  = \left\{ 
		%			\begin{array}{ll}
			%				q_i^{*}, & \text{if }u_i(t) = 1, \\
			%				1-q_i^{*}, &\text{otherwise}. 
			%			\end{array}\right. \label{theorem-sufficient-condition-u}
		%	\end{align} 
	%	Substituting (\ref{theorem-sufficient-condition-u}) and $ \mathbb{P}(s_i(t) = 1) = p_i $ into (\ref{gamma-definition}), we can derive the distribution of $ \gamma_i $ under the randomized policy as follows: 	
	%	\begin{align}
		%		\mathbb{P}(\gamma_i(t))\left\{
		%		\begin{array}{ll}
			%			q_i^{*}p_i, &\text{if }\gamma_i(t) = 1, \\
			%			1-q_i^{*}p_i, &\text{otherwise}. 
			%		\end{array}\right. \label{theorem-sufficient-condition-gamma} 
		%	\end{align} 
	%	Considering (\ref{AoI}) and (\ref{theorem-sufficient-condition-gamma}), we can further conclude that under the randomized policy, $ \Delta_i $ is geometrically distributed, i.e., 
	%	\begin{align}
		%		\mathbb{P}(\Delta_i(t) = k) = q_i^{*}p_i(1-q_i^{*}p_i)^{k-1}. 
		%	\end{align}
	%	Thus, (\ref{stability-condition-2}) is a sufficient stability condition of system $ i $ under the randomized policy\cite[Theorem 12]{you2011mean}. 
	The proof is given in Appendix \ref{appendix-sufficient-condition-of-system-stability}. 
\end{IEEEproof} 

It is possible to obtain a tighter sufficient stability condition for the system using the proposed lightweight scheduling policy by deriving the AoI distribution. 
%Unfortunately, due to the complexity induced by the real-time sorting of the Whittle indexes, characterizing the AoI distribution of the system becomes complicated. 
We plan to investigate this in our future work.

\section{Simulation Study\label{section-simulation}}

In this section, we simulate the proposed lightweight scheduling policy and compare its performance with existing policies, including the AoI greedy policy\cite{kadota2018scheduling}, the VoI (Value of Information) greedy policy\cite{ayan2019age}, the AoI Whittle index-based policy\cite{kadota2018scheduling}, and the VoI Whittle index-based policy\cite{wang2019whittle}, in terms of the mean-square estimation error (MSE) and computation time. 
To highlight the difference between all the policies mentioned above and the optimal one, we also plot the theoretical lower bound (\ref{lower-bound}) in each figures. 

In our simulations, the dynamical systems as in (\ref{kalman-filter-state}) and (\ref{kalman-filter-observe}) are third-order plants with randomly generated matrices $ \{A_i, C_i, R_i, Q_i\} $, where controllability and observability conditions are ensured. 
%$ R_i $ and $ Q_i $ are diagonal positive-definite matrices. For each system $ i $, we choose the parameter $ \alpha_i $ to be $ \rho^2(A_i) $. In this case, the AoI Whittle index-based policy can be viewed as a special case of our lightweight scheduling policy if letting $ f_i(\Delta_i) = \alpha_i\Delta_i $. 
We set transmission probabilities $ \{p_i\} $ that satisfy the necessary stability condition (\ref{stability-condition-1}) for each system. Then we run the system independently $ 10^4 $ times and plot the average values in the following figures. 

\begin{figure}
	\centering
	\includegraphics[width = 0.48\textwidth]{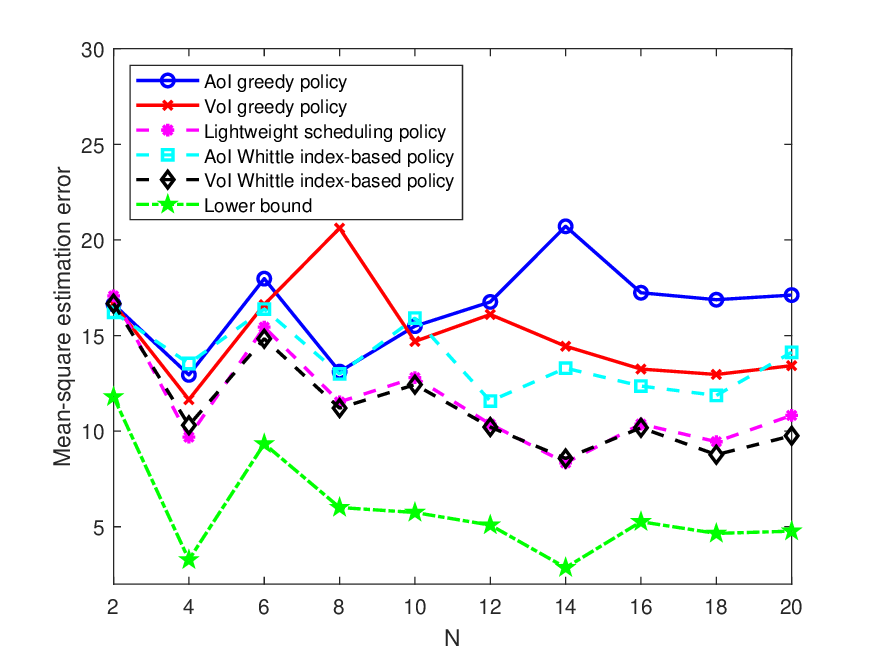}
	\caption{Performance comparison of the 
		five scheduling policies with $ N/M = 2 $. \label{figure_MN_2}}
\end{figure}

We first conduct simulations under different values of $ N $ and $ M $, with $ N/M $ fixed at 2. As shown in Fig. \ref{figure_MN_2}, our policy performs close to the VoI Whittle index-based policy while outperforms the others. 
This result is expected because the policies based on AoI operate regardless of the system dynamics and treat each system equally, leading to a performance loss. Conversely, the two greedy policies are ignorant of the successful transmission probabilities, which also results in a performance loss.  

%All policies perform similarly when $ M = 1, N = 2 $, which is reasonable because a policy can search a sub-optimal decision easily if the decision space is small. 

\begin{figure} 
	\centering
	\includegraphics[width = 0.48\textwidth]{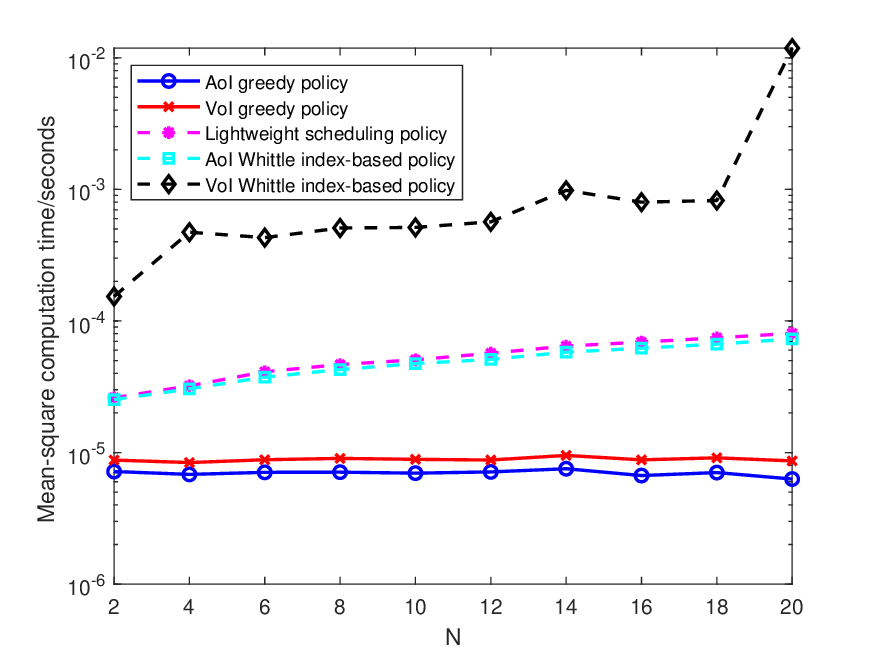} 
	\caption{Computation time comparison of the scheduling policies.  \label{figure_time}} 
\end{figure}

%\subsection{Computation Time Comparison}
To compare the computation time of different schedulers, we choose $N$ ranging from $1$ to $20$, with $M/N=2$ and $n_{\max} = 3$. Also, as far as we known, the VoI Whittle index is not in a close form, thus, during simulation, we compute the VoI Whittle index by iteration. 
As shown in Fig. \ref{figure_time}, regarding the computation time, the two greedy policies are the fastest, followed by our lightweight scheduling policy and the AoI Whittle index-based policy. The VoI Whittle index-based policy is the slowest and takes approximately $10$ times longer than our policy.
% Also, as the number of systems increases, the computation time of the greedy policies almost remains the same, while both the AoI Whittle index-based policy and our lightweight scheduling policy experience an increase in the computation time. However, the computation time of the VoI Whittle index-based policy grows faster than ours. 
These results match our expectations. As discussed in Section\ref{section-scheduling-policies}
%, both the proposed policy and the AoI Whittle index-based policy have an additional computational complexity of $ O(N) $ for computing the Whittle indexes. Nevertheless, 
the VoI Whittle index-based policy suffers from an extra complexity of $ O(Nn_{\max}^3) $ for matrix computations when compared with our policy.

\begin{figure}
	\centering
	\includegraphics[width = 0.48\textwidth]{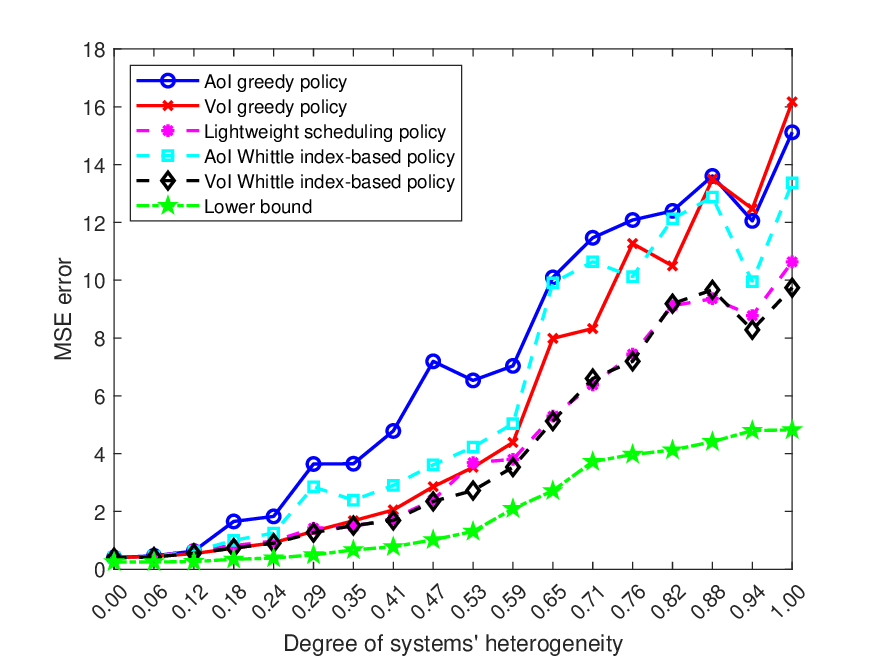}
	\caption{Performance comparison of the 
		scheduling policies under different degrees of systems heterogeneity.  \label{figure_homogeneous}}
\end{figure}

%\subsection{Impact of System Heterogeneity}
It is well known that if the dynamical systems are homogeneous, i.e., the sets of parameters $  \{A_i, C_i, R_i, Q_i, p_i\} $ are the same for different system $ i $, then all the policies are equivalent and degenerate into a Round-Robin sensor schedule. To evaluate how the heterogeneity  of systems impacts policy performance, we conduct further simulations with varying numbers of homogeneous systems, where the degree of heterogeneity is defined as the ratio of the number of systems that have distinct sets of parameters to the total number of systems. 
%Specifically, we initialize 18 homogeneous systems and gradually increase the degree of heterogeneity by replacing one of the remaining homogeneous systems with a randomly generated system distinct from all the others. 
%As shown in Fig. \ref{figure_homogeneous}, as the degree of heterogeneity increases, the performance of all five policies degrades. 
%In particular, when the degree of heterogeneity reduces near to 0, the five policies almost perform the same.
As shown in Fig. \ref{figure_homogeneous}, our policy performs similarly to the computationally expensive VoI Whittle index-based policy but significantly better than the other three policies in all cases. Furthermore, the performance gap between these policies widens as the degree of heterogeneity increases.

The systems' heterogeneity incurs a significant complexity in designing and analyzing scheduling policies for such multi-sensor remote estimation systems. There remains much to be discovered in the structure of the heterogeneity, of which the most intimate conclusion is that, as Fig. \ref{figure_homogeneous} shows, the lower the degree of heterogeneity is, the more suitable to treat the sensors similarly. 

	We present the system performance of the policies mentioned above and the optimal one obtained by Dynamic Programming (DP) in TABLE \ref{table-optimal} with randomly generated systems. It can be seen that for small scale of the system, the proposed policy performs close to the optimal one. Note that the state space increases exponentially as the number of systems increases, making it infeasible for larger-scale systems. 

\begin{table}[htbp] 
	\caption{Comparison of the MSE achieved by our lightweight scheduling policy with the optimal one. \label{table-optimal}}
	\centering
	\begin{tabular}{cccccc}
		\toprule
		M & 1 & 1 & 2  & 2 & 3 \\ 
		N & 2 & 3 & 3  & 4 & 4 \\ 
		\midrule 
		Ours & 5.55 & 5.55 & 4.13  & 17.6 & 6.58 \\ 
		Optimal & 5.34 & 5.32 & 4.00  & 15.3 & 6.55 \\ 
		\bottomrule
	\end{tabular}
\end{table}

\begin{figure}
	\centering
	\includegraphics[width = 0.48\textwidth]{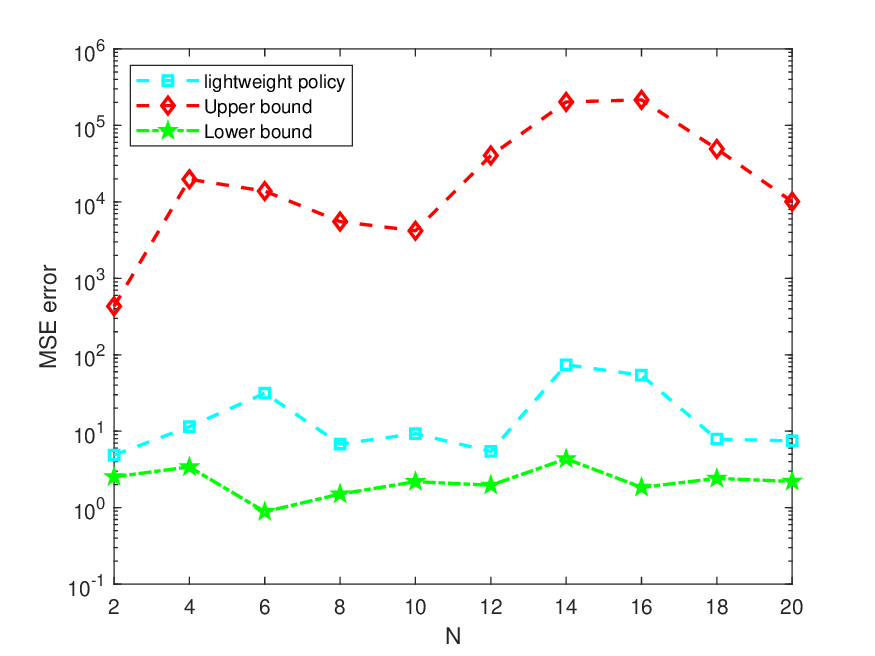}
	\caption{Performance comparison of the upper bound with the proposed policy under different system scales.  \label{fig-upperbound}}
\end{figure}

	Moreover, we present in Fig. \ref{fig-upperbound} the performance of the proposed lightweight policy and the upper bound. The results shown in Fig. \ref{fig-upperbound} indicate that the upper bound we have proposed is not tight. As the system scale becomes larger, the proposed upper bound becomes looser, which is as expected since the effect of the denominator of equation (\ref{upper-bound}) is getting larger. 

\begin{figure}
	\centering
	\includegraphics[width = 0.48\textwidth]{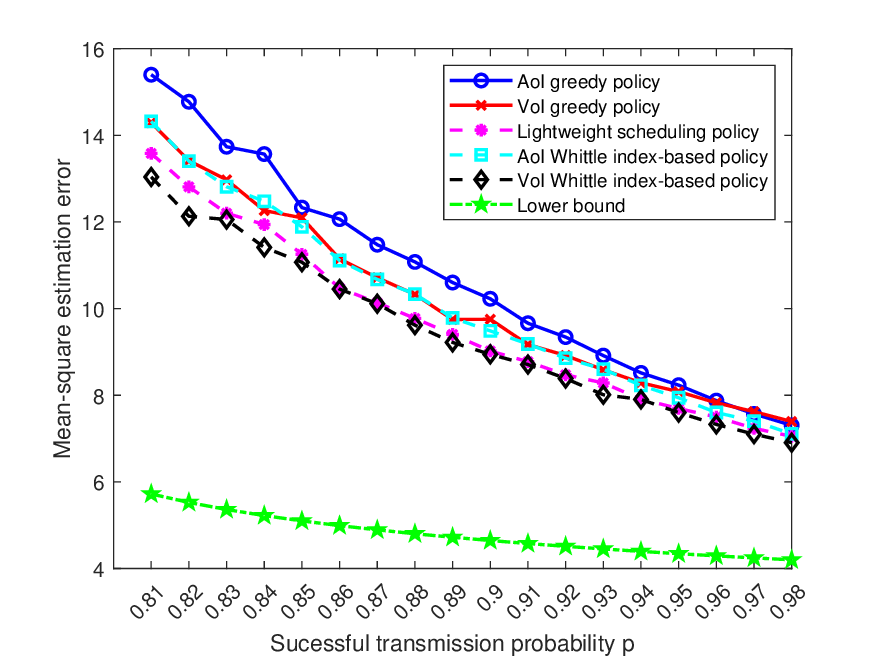}
	\caption{Performance comparison of the scheduling policies under different successful transmission probabilities.  \label{figure_p}}
\end{figure}

%\begin{figure}[htbp]
%	\centering
%	\subfloat[\tiny Performance comparison of the five scheduling policies with $ N/M = 2 $. ]{
	%		\label{figure_MN_2}
	%		\includegraphics[width = 0.24\textwidth]{files/figs/MN_2.eps}}\vspace{-0.1cm}%
%	\subfloat[\tiny Computation time comparison of the scheduling policies.]{
	%		\label{figure_time}
	%		\includegraphics[width = 0.24\textwidth]{files/figs/time_cost.eps}}\vspace{-0.1cm} 
%	
%	\subfloat[\tiny Performance comparison of the scheduling policies under different degrees of systems heterogeneity.]{
	%		\label{figure_homogeneous}
	%		\includegraphics[width = 0.24\textwidth]{files/figs/homogeneous.eps}}\vspace{-0.1cm}% 
%	\subfloat[\tiny Performance comparison of the scheduling policies under different successful transmission probabilities.]{
	%		\label{figure_p}
	%		\includegraphics[width = 0.24\textwidth]{files/figs/p.eps}}
%	\caption{Simulation results. }
%\end{figure}

%\subsection{Impact of Different Successful Transmission Probabilities}
Generally speaking, the larger the successful transmission probability is, the more accurate the estimator can be. We conduct more simulations to demonstrate the relationship between the mean-square estimation error achieved by the scheduling policies and the transmission probability. Considering the requirement of the system stability, for each system $ i $, we set the transmission probability $ p_i = p$ with $ p\in [0.8, 1]  $. As shown in Fig. \ref{figure_p}, there is a clear trend that the mean-square estimation error is decreasing as the successful transmission probability increases, which is as normally expected. We can further see that as $ p $ increases, the performance gap between these policies is getting smaller. The reason is that, as the successful transmission rate increases, intuitively, the influence of the previous scheduling decisions on the current one gets smaller. We also illustrate this phenomenon in view of the Bellman equation (\ref{decoupled Bellman_equation}). Since $ 1-p_i $, the transition probability from the previous state to the current one under decision $ u_i(t-1) = 1 $, is lower, the value function of the previous state is less influential on the current state, rendering that the performance of the optimal policy is close to the greedy one. 

\section{Conclusion\label{conclusion}}
This paper has studied the sensor scheduling problem of remote estimation systems under channel constraints. We have proposed a lightweight scheduling policy based on an AoI function built with the system characteristic parameters, and an upper and lower bound of the estimation performance under the proposed policy have been derived. Moreover, necessary and sufficient conditions for the remote estimation error stability have been established. Through simulations, we have demonstrated that our policy performs closely to the VoI Whittle index-based policy but is computationally much quicker. The simulation results have also clearly shown that the proposed policy outperforms the other three existing ones. 
%, and that the performance gap widens as the system heterogeneity increases or the successful transmission probability decreases. 
	In our future work, we will explore other lightweight policies in low-dimensional matrix form to achieve a better balance between computational complexity and scheduling performance. Additionally, we will incorporate more general network topologies.

\appendix 

%\subsection{Proof of Lemma \ref{lemma-trace-inequalities}\label{appendix-trace-inequalities}} 
%\begin{IEEEproof} 
%	Since that for any square matrices $ D $ and $ E $ of the same dimension, $ \textrm{Tr}(DE) = \textrm{Tr}(ED) $, \minew{
	%		we conclude that 
	%		\begin{align}
		%			\textrm{Tr}\left(A_i^kQ_i(A_i^T)^k\right) \nonumber
		%			= & \textrm{Tr}\left(A_i^TA_iA_i^{k-1}Q_i(A_i^T)^{k-1}\right)
		%		\end{align} 
	%		Considering that $ A_i^TA_i^k $ and $A_i^{k-1}Q_i(A_i^T)^{k-1}$ are all symmetric semi-definite matrices, we conclude that \cite{zhang2006eigenvalue}: 
	%		\begin{align}
		%			&\textrm{Tr}((A_i^k)^TA_i^kQ_i) \nonumber\\ 
		%			={} &\textrm{Tr}(A_i^TA_i^k)\textrm{Tr}(A_i^{k-1}Q_i(A_i^T)^{k-1}) \nonumber \\
		%			\le{} &\rho^{2}(A_i)\textrm{Tr}(A_i^{k-1}Q_i(A_i^T)^{k-1}) \nonumber\\
		%			\le{} &\rho^{2}(A_i)\rho^{2}(A_i)\textrm{Tr}(A_i^{k-2}Q_i(A_i^T)^{k-2}) \nonumber\\
		%			\le{} &\cdots 
		%			\le{} \rho^{2k}(A_i)\textrm{Tr}(Q_i).  \nonumber
		%		\end{align} 	
	%	}
%	In a similar argument, we can prove (\ref{trace-ineq-P}). 
%	
%	\minew{Moreover, as shown in \cite[Theorem 2]{zhang2006eigenvalue}, when $Q_i = U diag\{1, 0, \cdots\}U^{T}$, then $\textrm{Tr}((A_iA_i^T)^kQ_i) = \rho^{2k}(A_i) = \textrm{Tr}(A_i^kQ_i(A_i^T)^k)$. }
%\end{IEEEproof} 

\subsection{Proof of Theorem \ref{theorem decouple}\label{proof theorem decouple}} 
\begin{IEEEproof}
	First, we introduce a set of Lagrangian multipliers $ \{\mathcal{W}_i, i\in \mathcal{N}\} $ associated with the constraints of the decoupled problem (\ref{decoupled a}) and obtain the following problem: 
	\begin{align}
		\min_{\{\boldsymbol{u}\in \mathcal{U}\}}\quad & \lim_{\tau\rightarrow \infty}\frac{1}{\tau}\mathbb{E}
		\left[ \sum_{j=1}^\tau f_i(\Delta_i(j)) + \mathcal{W}_iu_i\right],  \forall i\in \mathcal{N}.\label{single_system_optimal_problem}
	\end{align}
	Note that the constant terms in the problem (\ref{decoupled_relaxed_optimal_problem}) have no impact on the optimal solution of the problem (\ref{single_system_optimal_problem}) and can be neglected. 
	For system $ i $, the Bellman equation of the problem (\ref{single_system_optimal_problem}) is
	\begin{align}
		V_i^d(\Delta_i)+\theta_i^d =
		\min_{u_i\in \{0, 1\}}\Big\{
		&\sum_{\Delta_i^{'}}
		\mathbb{P}(\Delta_i^{'}|\Delta_i, u_i) 
		V_i^d(\Delta_i^{'})\nonumber\\&+
		F_i(\Delta_i, u_i)\Big\}, 
		\label{decoupled Bellman_equation}
	\end{align}
	where $F_i(\Delta_i, u_i)\triangleq f_i(\Delta_i)+\mathcal{W}_iu_i$ denotes the loss function, $V_i^d(\Delta_i)$ denotes the value function, 
	and $\theta_i^d$ denotes the optimal cost. $ \mathbb{P}(\Delta_i^{'}|\Delta_i, u_i) $ denotes the conditional transition probability of the AoI $ \Delta_i $, which is given below based on  (\ref{AoI}):
	\begin{align}
		\mathbb{P}(\Delta_i^{'}|\Delta_i, u_i) = \left\{
		\begin{array}[]{ll}
			p_i, & \text{if }u_i = 1, \Delta_i^{'}=1,\\
			1-p_i, & \text{if }u_i = 1, \Delta_i^{'}=\Delta_i+1, \\
			1, &\text{if } u_i = 0, \Delta_i^{'}=\Delta_i+1, \\
			0, &\text{otherwise}. 
		\end{array} \right. \label{app-transition-probability}
	\end{align}
	Using the transition probability (\ref{app-transition-probability}), we simplify the Bellman equation (\ref{decoupled Bellman_equation}) as follows:
	\begin{align}
		V_i^d(\Delta_i)+\theta_i^d 
		% = \min_{u_i\in \{0, 1\}}\Big\{&		(1-p_i)u_iV_i^d(\Delta_i+1)+f_i(\Delta_i)\nonumber\\&+\mathcal{W}_iu_i+		(1-u_i)V_i^d(\Delta_i+1)\Big\}\nonumber\\
		= \min_{u_i\in \{0, 1\}}\Big\{&
		(1-p_iu_i)V_i^d(\Delta_i+1)\nonumber\\&+f_i(\Delta_i)+\mathcal{W}_iu_i\Big\}. 
		\label{simplified-Bellman-equation}
	\end{align}
	
	We divide the rest of this proof into two parts. The first part derives the expression of $ V_i^d(\Delta_i) $ and $ \theta_i^d $ by assuming the threshold structure (\ref{threshold-scheduler}), while the second part proves that this structure is the optimal one using the uniqueness property of the optimal solution of the Bellman equation. 
	
	\textbf{Part 1.} Assume that the optimal policy holds a threshold structure, i.e., equation (\ref{threshold-scheduler}) holds. Since for any $ \Delta_i < \Delta_{i, \text{th}} $, the optimal decision is $ u_i=0 $,  we conclude that the expected gain of decision $ u_i=1 $ in the Bellman equation (\ref{simplified-Bellman-equation}) is higher than that of $ u_i = 0 $, as the following inequality shows: 
	\begin{subequations}
		\begin{align}
			\mathcal{W}_i+(1-p_i)V_i^d(\Delta_i)&>{}V_i^d(\Delta_i),  \label{inequality-value-function-1} \\
			\mathcal{W}_i+(1-p_i)V_i^d(\Delta_i+1)&<{}V_i^d(\Delta_i+1),  \label{inequality-value-function-2}
		\end{align} %
	\end{subequations}
	where equation (\ref{inequality-value-function-2}) is derived for the case $ \Delta_i\ge \Delta_{i, \text{th}} $. 
	
	In the following, we determine the expressions of $ V_i^d(\Delta_i) $ and $ \theta_i^d $ for each system $ i $. 
	From (\ref{simplified-Bellman-equation}), we recursively obtain that: if $ \Delta_i \le \Delta_{i, \text{th}} $, then 
	\begin{align}
		V_i^d(\Delta_{i, \text{th}}-\tilde{\Delta}_i)={}&\sum_{k=1}^{\tilde{\Delta}_i}f_i(\Delta_{i, \text{th}}-k)+V_i^d(\Delta_{i, \text{th}})-\tilde{\Delta}_i\theta_i^d, \label{equality V1}
	\end{align}
	where $ \tilde{\Delta}_i = \Delta_{i, \text{th}} - \Delta_i $. 
	Similarly, if $ \Delta_i > \Delta_{i, \text{th}}, $ then 
	\begin{align}
		V_i^d(\Delta_i)={}&f_i(\Delta_i)+(1-p_i)V_i^d(\Delta_i+1)+\mathcal{W}_i-\theta_i^d.  \nonumber
	\end{align}
	Assuming that 
	\begin{align} 
		\lim_{j\rightarrow\infty}(1-p_i)^{j+1}V_i^d(\Delta_i+j+1)=0, \quad \forall i\in \mathcal{N} ,  \label{assumption V}
	\end{align}
	we have 
	\begin{align}
		V_i^d(\Delta_i)={}&\lim_{j\rightarrow\infty}\Big\{
		(1-p_i)^{j+1}V_i^d(\Delta_i+j+1)\nonumber\\&+\sum_{k=0}^j(1-p_i)^kf_i(\Delta_i+k)\nonumber\\&+
		(\mathcal{W}_i-\theta_i^d)\sum_{k=0}^j(1-p_i)^k\Big\}\nonumber \\
		={} &\sum_{k=0}^\infty(1-p_i)^kf_i(\Delta_i+k)+\frac{\mathcal{W}_i-\theta_i^d}{p_i}\nonumber\\
		={} &\frac{\beta_i\alpha_i^{\Delta_i}}{1-\alpha_i+p_i\alpha_i} + \frac{\mathcal{W}_i-\theta_i^d}{p_i}\label{equality V2}.
	\end{align} 
	Combining (\ref{equality V1}) and (\ref{equality V2}), we obtain the following expression of $ V_i^d(\Delta_i) $: 
	\begin{align}
		&V_i^d(\Delta_i)\nonumber\\=&\left\{
		\begin{array}{ll}
			\frac{\beta_i(\alpha_i^{\Delta_{i, \text{th}}} - \alpha_i^{\Delta_i})}{\alpha_i-1} + \Delta_i\theta_i^d + V_{i, \text{res}}, 
			&\text{if } \Delta_i < \Delta_{i, \text{th}},  \\ 
			\frac{\beta_i\alpha_i^{\Delta_i}}{1-\alpha_i+p_i\alpha_i} + \frac{\mathcal{W}_i-\theta_i^d}{p_i}, &\text{otherwise}, 
		\end{array}
		\right.\label{V}
	\end{align}
	where $ V_{i, \text{res}} \triangleq V_i^d(\Delta_{i, \text{th}})-\Delta_{i, \text{th}}\theta_i^d  $.  
	
	The fact that  the assumption (\ref{assumption V}) is consistent with (\ref{V}) can be easily verified as follows: 
	\begin{align}
		&\lim_{j\rightarrow\infty}(1-p_i)^{j+1}V_i^d(\Delta_i+j+1)\nonumber\\
		%		={} &\lim_{j\rightarrow\infty}\left\{		\sum_{k=0}^\infty(1-p_i)^{k+j+1}f_i(\Delta_i+k)\right.\nonumber\\&+		\left.\frac{\mathcal{W}_i-\theta_i^d}{p_i}(1-p_i)^{j+1}\right\}\nonumber\\
		={} &\lim_{j\rightarrow\infty}
		\left\{\sum_{k=0}^\infty(1-p_i)^{k+j+1}\beta_i\alpha_i^{\Delta_i +k}\right.\nonumber\\&\left.+\frac{\mathcal{W}_i-\theta_i^d}{p_i}(1-p_i)^{j+1}\right\}\nonumber\\
		\overset{(a)}{=}{} &\lim_{j\rightarrow\infty}
		\left\{(1-p_i)^{j+1}\left[\frac{\beta_i\alpha_i^{\Delta_i}}{1-\alpha_i(1-p_i)}+\frac{\mathcal{W}_i-\theta_i^d}{p_i}\right]\right\}, \nonumber
		%		\label{app-equality-V}
	\end{align}
	where (a) is due to the assumption that $ \alpha_i(1-p_i) < 0 $ (which is also the necessary stability condition as Theorem \ref{theorem-necessary-condition-of-system-stability} shows). 
	
	To obtain the expression of $ \theta_i^d $, we first assign $ V_i(1)=0 $ and obtain the following equation based on (\ref{V}) as follows: 
	\begin{align} 
		\frac{\beta_i(\alpha_i^{\Delta_{i, \text{th}}}-\alpha_i)}{1-\alpha_i}+V_i^d(\Delta_{i, \text{th}})-
		(\Delta_{i, \text{th}}-1)\theta_i^d=0. \label{V(1) = 0}
	\end{align} 
	Again, from (\ref{V}), we obtain $ V_i^d(\Delta_{i, \text{th}}) $ as follows: 
	\begin{align}
		V_i^d(\Delta_{i, \text{th}}) = \frac{\beta_i\alpha_i^{\Delta_{i, \text{th}}}}{1-\alpha_i+p_i\alpha_i}+\frac{\mathcal{W}_i-\theta_i^d}{p_i}. \label{V it}
	\end{align}
	By substituting (\ref{V it}) into (\ref{V(1) = 0}) and performing some simple deductions, we obtain the expression of $ \theta_i^d $ as follows: 
	\begin{align} 
		\theta_i^d = {}&\frac{\mathcal{W}_i+[p_i\beta_i(\alpha_i^{\Delta_{i, \text{th}}}-\alpha_i)]/(1-\alpha_i)}{1+p_i\Delta_{i, \text{th}}-p_i}\nonumber\\&
		+{}\frac{p_i\beta_i\alpha_i^{\Delta_{i, \text{th}}}/(1-\alpha_i+p_i\alpha_i)}{1+p_i\Delta_{i, \text{th}}-p_i}. \label{theta}
	\end{align} 
	
	\textbf{Part 2.} Then we need to prove that the threshold structure assumption is consistent with (\ref{V}). First, we show that $ V_i^d(\Delta_i) $ is monotonically increasing under the assumptions $ \alpha_i > 1 $ and $ \alpha_i(1-p_i) < 1 $. 
	
	Next, we need to prove that the threshold structure assumption is consistent with (\ref{V}). First, we will demonstrate that $ V_i^d(\Delta_i) $ is monotonically increasing under the assumptions $ \alpha_i > 1 $ and $ \alpha_i(1-p_i) < 1 $.
	We write the difference equation of $ V_i^d(\Delta_i) $ as follows: 
	\begin{align}
		&V_i^d(\Delta_i+1)-V_i^d(\Delta_i)\nonumber\\=&\left\{
		\begin{array}{ll}
			\theta_i - \beta_i\alpha_i^{\Delta_i}, &\text{if }\Delta_i < \Delta_{i, \text{th}},  \\
			\frac{\beta_i\alpha_i^{\Delta_i}(\alpha_i-1)}{1-\alpha_i+p_i\alpha_i}, 
			&\text{otherwise}.
			\label{V-minus}
		\end{array}
		\right.
	\end{align}
	Simplifying (\ref{inequality-value-function-2}) at $ \Delta_i=\Delta_{i, \text{th}} $,  we derive the following inequality: 
	\begin{align} 
		\mathcal{W}_i\le p_iV_i^d(\Delta_{i, \text{th}}+1). \label{app-W-V1}
	\end{align} 
	Considering that $ \Delta_{i, \text{th}} $ is the critical threshold, we can further conclude that the following inequality holds (otherwise the threshold should be $ \Delta_{i, \text{th}} - 1 $): 
	\begin{align} 
		\mathcal{W}_i\ge p_iV_i^d(\Delta_{i, \text{th}}).  \label{app-W-V2}
	\end{align} 
	By substituting (\ref{V}) into (\ref{app-W-V2}) and performing some simplifications, we obtain the following inequality: 
	\begin{align}
		\theta_i^d \ge{} &p_i\frac{\beta_i\alpha_i^{\Delta_{i, \text{th}}}}{1-\alpha_i+p_i\alpha_i}\nonumber\\
		\overset{(a)}{\ge}{} & \beta_i\alpha_i^{\Delta_{i, \text{th}}}\nonumber\\
		\overset{(b)}{\ge}{} & \beta_i\alpha_i^{\Delta_i}, \quad \forall \Delta_i < \Delta_{i, \text{th}}, \label{inequality difference equation V 1}
	\end{align}
	where (a) and (b) hold since $ \alpha_i > 1 $ and $ \alpha_i(1-p_i) < 1 $. Similarly, we conclude that
	\begin{align}
		\frac{\beta_i\alpha_i^{\Delta_i}(\alpha_i-1)}{1-\alpha_i+p_i\alpha_i} > 0.\label{inequality difference equation V 2}
	\end{align}
	
	By combining (\ref{inequality difference equation V 1}), (\ref{inequality difference equation V 2}) and (\ref{V-minus}), we can prove that $ V_i^d(\Delta_i+1)-V_i^d(\Delta_i)>0 $, i.e., $ V_i^d(\Delta_i) $ is monotonically increasing in $ \Delta_i $. This indicates that the threshold structure assumption is consistent with (\ref{V}). 
	Considering the well-known conclusion that the unique solution of the Bellman equation is its optimal solution, we further conclude that the assumed threshold-structured solution is the optimal one of the Bellman equation (\ref{decoupled Bellman_equation}). Thus, we complete the proof. 
\end{IEEEproof}

\subsection{Proof of Theorem \ref{theorem whittle index}\label{appendix whittle index}}
\begin{IEEEproof}
	Remember that the Whittle index is defined as the critical value of the Lagrangian multiplier at which both decisions $ u_i = 0 $ and $ u_i = 1 $ yield the same expected value. 
	By substituting (\ref{V}) into (\ref{app-W-V1}) and performing some simplification, we obtain
	\begin{align}
		\theta_i^d\le \frac{p_i\beta_i\alpha_i^{\Delta_{i, \text{th}}+1}}{1-\alpha_i+p_i\alpha_i}. \label{app-W-ineq-2}
	\end{align}
	Substituting (\ref{theta}) into (\ref{app-W-ineq-2}), we conclude that the following inequality holds after some simplification: 
	\begin{align}
		\mathcal{W}_i \le{} &\frac{\beta_ip_i^2\Delta_{i, \text{th}}\alpha_i^{\Delta_{i, \text{th}} + 1}}{1+\alpha_ip_i - \alpha_i} - \frac{\beta_ip_i\alpha_i(\alpha_i^{\Delta_{i, \text{th}}} - 1)}{\alpha_i - 1}. \label{app-W}
	\end{align}
	By replacing $ \Delta_{i, \text{th}} $ with $ \Delta_i $ and simplifying the RHS of (\ref{app-W}), we obtain the Whittle index in (\ref{W}). 
	
	To verify the indexability of the Whittle index policy, we express the passive set (the set of $ \Delta_i $ for which the decision $ u_i = 0 $ is the optimal one) of the problem (\ref{simplified-Bellman-equation}) as follows: 
	\begin{align}
		\mathcal{P}_i = \{\Delta_i|\Delta_{i}\in \mathbb{N}^+, \mathcal{W}_i+(1-p_i)V_i(\Delta_i)>{}V_i(\Delta_i)\}. \nonumber
	\end{align}
	In addition, by examining the derivative of the Whittle index (\ref{W}), we observe that it monotonically increases with the AoI $ \Delta_i $. Thus, as $ \Delta_i $ increases from $ 1 $ to $ \infty $, the Whittle index increases from $ \mathcal{W}_i(1) $ to $ \lim_{\Delta_i\rightarrow \infty}\mathcal{W}_i(\Delta_i) \rightarrow \infty $ and the set $ \mathcal{P}_i $ expands from $ \emptyset $ to $ \mathbb{N}^+ $, thereby proving the indexability. 
\end{IEEEproof} 

\subsection{Proof of Theorem \ref{theorem-lower-bound}\label{appendix-lower-bound}} 
\begin{IEEEproof}
	Applying (\ref{dmdp-stable-distribution}) to (\ref{relaxed_optimal_problem a}) yields 
	\begin{align}
		&\lim_{\tau\rightarrow \infty} \frac{1}{\tau}\mathbb{E}\left[ \sum_{t=1}^\tau\sum_{i=1}^N u_i(t)\right]\nonumber\\
		= {}&\sum_{i = 1}^N\left(1-\sum_{\Delta_i = 1}^{\Delta_{i, \text{th}} - 1} \Psi_i(\Delta_i)\right)  \nonumber\\
		={} &\sum_{i = 1}^N\left(1-\sum_{\Delta_i = 1}^{\Delta_{i, \text{th}} - 1} \frac{p_i}{\Delta_{i, \text{th}}p_i+1-p_i}\right) \nonumber\\
		={} &\sum_{i = 1}^N\frac{1}{\Delta_{i, \text{th}}p_i+1-p_i}\le M. \nonumber
	\end{align}
	We can reformulate the objective function $ J $ similarly and obtain the optimal problem (\ref{dmdp-optimal-problem}). 
	For any given thresholds $ \{\Delta_{i, \text{th}}, i\in \mathcal{N}\} $, the performance $ J_{relaxed} $ can be calculated as follows: 
	\begin{align}
		J_{relaxed} = {}&\sum_{i = 1}^N\sum_{\Delta_i=1}^\infty f_i(\Delta_i)\Psi_i(\Delta_i)\nonumber\\
		= {}&\sum_{i = 1}^N\sum_{\Delta_i=1}^{\Delta_{i, \text{th}}} f_i(\Delta_i)\Psi_i(\Delta_i)\nonumber\\&+\sum_{i = 1}^N\sum_{\Delta_i=\Delta_{i, \text{th}}+1}^{\infty} f_i(\Delta_i)\Psi_i(\Delta_i)\nonumber\\
		= {}&\sum_{i = 1}^N\sum_{\Delta_i=1}^{\Delta_{i, \text{th}}} \frac{p_i\beta_i\alpha_i^{\Delta_i}}{\Delta_{i, \text{th}}p_i+1-p_i}\nonumber\\&+\sum_{i = 1}^N\sum_{\Delta_i=\Delta_{i, \text{th}}+1}^{\infty} \frac{p_i\beta_i\alpha_i^{\Delta_i}(1-p_i)^{\Delta_i-\Delta_{i, \text{th}}}}{\Delta_{i, \text{th}}p_i+1-p_i}. \label{app-J-relaxed} 
	\end{align}
	Since we have assumed that $ \alpha_i(1-p_i) < 1 $, we conclude that 
	\begin{align}
		\sum_{i = 1}^N\sum_{\Delta_i=\Delta_{i, \text{th}}+1}^{\infty} \frac{p_i\beta_i\alpha_i^{\Delta_i}(1-p_i)^{\Delta_i-\Delta_{i, \text{th}}}}{\Delta_{i, \text{th}}p_i+1-p_i} < \infty. \nonumber
	\end{align}
	By simplifying (\ref{app-J-relaxed}), we obtain the following equation: 
	\begin{align}
		J_{relaxed}	={}& \sum_{i = 1}^N\frac{p_i\beta_i}{\Delta_{i, \text{th}}p_i+1-p_i}\left[\frac{\alpha_i(1-\alpha_i^{\Delta_{i, \text{th}}})}{1-\alpha_i}\right.\nonumber\\&\left. + \frac{\alpha_i^{\Delta_{i, \text{th}}+1}(1-p_i)}{1-\alpha_i(1-p_i)}\right]. \nonumber
	\end{align}
	Thus, the equation (\ref{lower-bound}) can be deduced.
	
	Revisiting Theorem \ref{theorem decouple}, we observe that the choice of $ \{\alpha_i\} $ and $ \{\beta_i\} $ is unrelated to the threshold structure. We further conclude that the optimal performance of the optimization problem with the parameters $ \{\hat{\alpha}_i\} $ and $ \{\hat{\beta}_i\} $ satisfies the following inequalities
	\begin{align} 
		\textrm{Tr}(P_i(t))\ge \hat{f}(\Delta_{i}) = \hat{\beta}_i\hat{\alpha}_i^{\Delta_i},  \nonumber
		%		\label{app-inequality-f}
	\end{align} 
	is $\underline{J}_{origin}$. To specify $ \{\hat{\alpha}_i\} $ and $ \{\hat{\beta}_i\} $, we consider the following inequalities: 
	\begin{align} 
		\textrm{Tr}(A_i^kQ_i(A_i^k)^T)&\overset{(a)}{\ge}  \lambda_{\min}(Q_i)\textrm{Tr}(A_i^k(A_i^k)^T) \nonumber\\
		&\overset{(b)}{\ge}\lambda_{\min}(Q_i)\zeta_i\rho(A_i)^{2k}, \nonumber \\
		\textrm{Tr}(A_i^k\bar{P}_i(A_i^k)^T)&\ge\lambda_{\min}(\bar{P}_i)\zeta_i\rho(A_i)^{2k}. \nonumber 
	\end{align}
	One can refer to \cite{fang1994inequalities} for the proof of (a) and to \cite[Lemma 3]{cao2014cognitive} for the proof of (b). 
	Therefore, we choose $ \hat{\beta}_i = \zeta_i\min\{\lambda_{\min}(\bar{P}_i), \lambda_{\min}(Q_i)\} $ and $ \hat{\alpha}_i = \rho^2(A_i) $, and then obtain the expression of $ \underline{J}_{origin} $ by replacing $ \{\alpha_i\} $ and $ \{\beta_i\} $ in (\ref{lower-bound}) with $ \{\hat{\alpha}_i\} $ and $ \{\hat{\beta}_i\} $, respectively. 
\end{IEEEproof}

\subsection{Proof of Lemma \ref{lemma-searching-set}\label{appendix-searching-set}}
\begin{IEEEproof} 
	We first write the boundary of the searching space of  (\ref{dmdp-optimal-problem-b}) as $ \mathcal{B}\triangleq \{\{\Delta_{i, \text{th}}^{\mathcal{B}}|i\in \mathcal{N}\}\} $, where
	\begin{align}
		\sum_{i\in \mathcal{N}}\frac{1}{\Delta_{i, \text{th}}^{\mathcal{B}}p_i+1-p_i} &\le M, \label{lemma2-ineq-1}
	\end{align} 
	and for any $ i\in \mathcal{N} $, 
	\begin{align}
		\sum_{j\in \mathcal{N}, j\neq i}\frac{1}{\Delta_{j, \text{th}}^{\mathcal{B}}p_j+1-p_j} + \frac{1}{\Delta_{i,  \text{th}}^{\mathcal{B}}p_i+1-2p_i} &> M. \label{lemma2-ineq-2}
	\end{align} 
	
	Next, we prove that the optimal thresholds $ \{\Delta_{i, \text{th}}^{*}\} $ belong to $\mathcal{B} $ by contradiction. First, we assume that $ \{\Delta_{i, \text{th}}^{*}\} \notin \mathcal{B} $. Since $ \{\Delta_{i, \text{th}}^{*}\} $ satisfies (\ref{lemma2-ineq-1}), we can find $ \hat{\Delta}_{i, \text{th}}\in \mathcal{B} $ that satisfies the following inequality by decreasing the elements of $ \{\Delta_{i, \text{th}}^{*}\} $: 
	\begin{align} 
		\hat{\Delta}_{i, \text{th}} &\le \Delta_{i, \text{th}}^{*}, \quad \forall i\in \mathcal{N}. \label{app-lemma2-assumption-1}
	\end{align} 
	The optimality of  $ \{\Delta_{i, \text{th}}^{*}\} $ indicates that 
	\begin{align}
		\hat{J}_{relaxed} &\ge J^{*}_{relaxed}. \label{app-lemma2-assumption-2}
	\end{align}
	However, by calculating $ \frac{\partial J_{relaxed}}{\partial \Delta_{i, \text{th}}} $ as follows: 
	\begin{align}
		\frac{\partial J_{relaxed}}{\partial \Delta_{i, \text{th}}} = &\frac{\mathcal{K}_i}{\mathcal{D}_i} \alpha_i^{\Delta_{i, \text{th}}}\left(p_i^2\log(\alpha_i)\Delta_{i, \text{th}}+\mathcal{R}_i\right) \nonumber\\&+ p_i(1-\alpha_i+\alpha_ip_i), \nonumber
	\end{align}
	where $ \mathcal{D}_i=(1-p_i+p_i\Delta_{i, \text{th}})^2>0 $, $ \mathcal{R}_i = (1-p_i)p_i\log(\alpha_i)-p_i^2 $, $ p_i^2\log(\alpha_i)>0 $, $ p_i(1-\alpha_i+\alpha_ip_i)>0 $ and $ \frac{\mathcal{K}_i}{\mathcal{D}_i} \alpha_i^{\Delta_{i, \text{th}}}>0 $, we conclude that $ \frac{\partial J_{relaxed}}{\partial \Delta_{i, \text{th}}} $ has only one zero point, denoted by $ \Delta_i^{0} $, and $ J_{relaxed} $ is decreasing when $\Delta_i \in [0, \Delta_i^{0}]$ while increasing when $\Delta_i\in (\Delta_i^{0}, \infty] $.Furthermore, we notice that
	\begin{align}
		\mathcal{K}_i\frac{p_i\alpha_i^2-\alpha_ip_i+\alpha_i-1}{1+p_i}>	\mathcal{K}_i\frac{\alpha_i-1}{1}, \nonumber
	\end{align}
	i.e., the $ i $'th element of the summation in $ J_{relaxed} $ under the threshold $ \Delta_{i, \text{th}}=1 $ is smaller than that under the threshold $ \Delta_{i, \text{th}}=2 $. This indicates that
	\begin{align} 
		\Delta_{i}^{0} < 2. \nonumber
	\end{align} 
	Thus, $ J_{relaxed} $ increases when $\Delta_i\in [2, \infty) $. Considering that $ \Delta_{i, \text{th}} \in \mathbb{N}^{+} $, we conclude that  $ J_{relaxed} $ increases with $ \Delta_{i, \text{th}} $. This results in a contradiction between (\ref{app-lemma2-assumption-1}) and (\ref{app-lemma2-assumption-2}). Hence, we have proven that $ \{\Delta_{i, \text{th}}^{*}\} \in\mathcal{B} $. 
	
	To prove (\ref{lemma-upperbound}), we focus on an arbitrarily chosen $ i $. Considering (\ref{lemma2-ineq-2}), we notice that the term $ \frac{1}{\Delta_{i, \text{th}}^{*}p_i+1-2p_i} $ is lower bounded by the smallest positive gap as follows: 
	\begin{align} 
		\frac{1}{\Delta_{i, \text{th}}^{*}p_i+1-2p_i}\ge \mathcal{G}_{i, \min}. \nonumber
	\end{align} 
	Rearranging the above inequality, we conclude that (\ref{lemma-upperbound}) holds, thereby completing the proof. 
\end{IEEEproof} 
	\subsection{Proof of Lemma \ref{lemma-lyapunov-drift-inequality}\label{appendix-lyapunov-drift-inequality}} 
	\begin{IEEEproof} 
		Given that the optimal policy that minimizes (\ref{lyapunov-drift}) is the greedy policy proposed by Kim \emph{et al.}\cite{kim2014scheduling}, the Lyapunov drift obtained based on the following randomized policy should serve as an upper bound for (\ref{lyapunov-drift}). 
		%The randomized policy in this paper schedules the sensors as follows. 
		Considering the constraint (\ref{optimal-problem-b}), there are a total of $ \mathbb{C}^M_N $ choices of scheduling decisions, denoted by $ \{{\mathbf u}_j\in \mathcal{U}, j\in\{1,..., \mathbb{C}^M_N\}\} $. Thus, we schedule the sensors by randomly selecting one of the choices with probability $ q^c_j=\mathbb{P}({\mathbf u}_j),  \forall j\in \{1, \cdots, \mathbb{C}^M_N\} $, where $ \sum q^c_j=1 $. This results in that the randomized policy schedules each sensor with a certain probability calculated as follows: 
		\begin{align} 
			q_i=\sum_{j:u_{j,i}=1}q^c_j, \quad \forall i\in \mathcal{N}, \label{randomized-policy-q}
		\end{align}
		where $ u_{j,i} $ is the $ i $th element of $ {\mathbf u}_j $. Conversely, given $ \{q_i\} $, we can certainly find a feasible solution for the randomized policy $ \{p^c_j\} $ by solving (\ref{randomized-policy-q}), since the number of variables is fewer than that of equations.
		
		We denote $ \mathbb{E}[\gamma_i(t)|\Delta_i(t)] $ obtained from the randomized policy by $ \mathbb{E}[\gamma_i^r(t)] $. Considering that the conditional expectation is larger, i.e., $ \mathbb{E}[\gamma_i(t)|\Delta_i(t)] > \mathbb{E}[\gamma_i^r(t)] $, we can derive the following inequality: 
		\begin{align}
			\vec{\mathcal{L}}(t) \le{} & \sum_{i = 1}^N -\mathbb{E}[\gamma_i^r(t)]\Big[l_{i1}\alpha_i\Delta_i\alpha_i^{\Delta_i} + (l_{i1} + l_{i2})
			\alpha_i\alpha_i^{\Delta_i} \nonumber\\&- (l_{i1} + l_{i2})\alpha_i\Big] 
			+\Big[l_{i1}(\alpha_i - 1) \Delta_{i}\alpha_i^{\Delta_i} \nonumber\\&+ (l_{i1}\alpha_i+l_{i2}\alpha_i - l_{i2})\alpha_i^{\Delta_i}\Big]. \nonumber 
		\end{align}	
		
		In the following, we introduce an optimization problem to obtain $ \mathbb{E}[\gamma_i^r(t)] $. 
		Since the randomized policy schedules sensor $ i\in \mathcal{N} $ with probability $p_i$, where $ \sum_{i = 1}^Nq_i \le M, q_i\in(0, 1) $, we consider the process of successful transmission as an arithmetic renewal process with a unit span $ \{X_n; n\ge1\} $. 
		As $ t\rightarrow \infty, $ $ \Delta_i(t)$ follows the following distribution\cite[Theorem 5.7.1]{gallager2013stochastic}: 
		\begin{align}
			\lim_{t\rightarrow \infty}\mathbb{P}_{\Delta_i(t)}(n) = \frac{\sum_{m=n}^\infty \mathbb{P}_{X_i}(m)}{\mathbb{E}[X_i]}. \nonumber
		\end{align}
		We immediately have the following equation
		\begin{align}
			\mathbb{E}\left[\alpha_i^{\Delta_i(t)}\right] 
			& = \frac{1}{\mathbb{E}[X_i]}\sum_{n = 1}^\infty\sum_{m = n}^\infty\alpha_i^{n}\mathbb{P}_{X_i}(m)\nonumber\\
			& = \frac{1}{\mathbb{E}[X_i]}\sum_{m = 1}^\infty\sum_{n = 1}^m\alpha_i^{n}\mathbb{P}_{X_i}(m)\nonumber\\
			& = \frac{1}{\mathbb{E}[X_i]}\sum_{m = 1}^\infty\alpha_i\frac{1-\alpha_i^{m}}{1-\alpha_i}\mathbb{P}_{X_i}(m)\nonumber\\
			& = \frac{\alpha_i}{(1-\alpha_i)\mathbb{E}[X_i]}(1 - \mathbb{E}[\alpha_i^{X_i(t)}]).\label{delta_X}
		\end{align}
		Considering the definition of $ \{X_n\} $, it can be obtained that $ \forall i\in \mathcal{N} $, 
		\begin{align}
			\mathbb{E}[\gamma_i^r(t)] &= {}p_iq_i,  \nonumber \\ % \label{equation-E-gamma}
			\mathbb{P}_{X_i}(n) &= {}q_ip_i(1-q_ip_i)^{n-1}.\label{PX} 
		\end{align}
		We calculate $ \mathbb{E}\left[\alpha_i^{X_i(t)}\right] $ as follows: 
		\begin{align}
			\mathbb{E}\left[\alpha_i^{X_i(t)}\right] = {}&\sum_{n = 1}^\infty \alpha_i^n\mathbb{P}_{X_i}(n)\nonumber\\
			= {}&\sum_{n = 1}^\infty \alpha_i^nq_ip_i(1-q_ip_i)^{n-1}\nonumber\\
			\overset{(a)}{=}{} &\frac{\alpha_iq_ip_i}{1-\alpha_i(1-p_iq_i)}, \label{equation-E-alpha-x}
		\end{align}
		where (a) is a result of the condition (\ref{problem-statical-scheduler-c}). Similarly, we can compute $ \mathbb{E}\left[X_i(t)\right] $ by
		\begin{align}
			\mathbb{E}\left[X_i(t)\right] = \frac{1}{p_iq_i}.\label{equation-E-x}
		\end{align}
		Substituting (\ref{equation-E-alpha-x}) and (\ref{equation-E-x}) into (\ref{delta_X}) yields 
		\begin{align} 
			\mathbb{E}\left[\alpha_i^{\Delta_i}\right] &= \frac{\alpha_ip_iq_i}{1-\alpha_i}(1-\frac{\alpha_ip_iq_i}{1-\alpha_i(1-p_iq_i)})\nonumber\\
			&= \frac{\alpha_ip_iq_i}{1-\alpha_i+\alpha_ip_iq_i}. \nonumber
			%		\label{equation-E-alpha}
		\end{align} 
		Thus, 
		\begin{align}
			\sum_{i = 1}^N\mathbb{E}\left[f_i(\Delta_i)\right] = &\sum_{i = 1}^N\mathbb{E}\left[\beta_i\alpha_i^{\Delta_i}\right] \nonumber\\
			=& \sum_{i = 1}^N\frac{\beta_i\alpha_ip_iq_i}{1-\alpha_i+\alpha_ip_iq_i}. 
			%		\label{equation-Ef} 
		\end{align}
		Therefore, we can reformulate problem (\ref{optimal_problem}) under the randomized policy as (\ref{problem-statical-scheduler-a}) 
		%	\begin{align}
			%		\min_{\{\boldsymbol{q}\in \mathcal{Q}\}}\quad &\sum_{i = 1}^N\frac{\beta_i\alpha_ip_iq_i}{1-\alpha_i+\alpha_ip_iq_i}, \label{problem-statical-scheduler-a} 
			%	\end{align}
		under the constraints (\ref{problem-statical-scheduler-b}), (\ref{problem-statical-scheduler-c}) and (\ref{problem-statical-scheduler-c}). By rearranging (\ref{problem-statical-scheduler-a}) and discarding the constant terms, we obtain the problem (\ref{problem-statical-scheduler}), thus completing the proof. 
	\end{IEEEproof}
	
	\subsection{Proof of Theorem \ref{theorem-upper-bound}\label{appendix-upper-bound}} 
	By rewriting (\ref{AoI}) as the following form: 
	\begin{align}
		\Delta_{i}(t+1) = \gamma_i(t) + (1-\gamma_i(t))(\Delta_i(t) + 1), \quad \forall i\in \mathcal {N},   \nonumber
		%	\label{equation-delta-i}
	\end{align} 
	we can reformulate (\ref{definition-lyapunov-drift}) as follows: 
	\begin{align}
		\vec{\mathcal{L}}(t)= & \sum_{i = 1}^N \Big\{(1-\mathbb{E}[\gamma_i(t)|\Delta_i(t)])\left[\mathcal{L}_i(\Delta_i(t) + 1) \right.\nonumber\\&\left.- \mathcal{L}_i(\Delta_i(t))\right] + \mathbb{E}[\gamma_i(t)|\Delta_i(t)][\mathcal{L}_i(1) - \mathcal{L}_i(\Delta_i(t))]\Big\}\nonumber\\
		=&  \sum_{i = 1}^N \big[-\mathbb{E}[\gamma_i(t)|\Delta_i(t)]\left(\mathcal{L}_i(\Delta_i(t) + 1) - \mathcal{L}_i(1)\right) \nonumber\\&+ \mathcal{L}_i(\Delta_i(t) + 1) - \mathcal{L}_i(\Delta_i(t))\big]\nonumber\\
		=&  \sum_{i = 1}^N -\mathbb{E}[\gamma_i(t)|\Delta_i(t)]\Big\{l_{i1}\alpha_i\Delta_i(t)\beta_i\alpha_i^{\Delta_i(t)} \nonumber\\& - \alpha_i(l_{i1} + l_{i2}) + \alpha_i(l_{i1} + l_{i2})
		\beta_i\alpha_i^{\Delta_i(t)}\Big\} 
		\nonumber\\& +\Big\{l_{i1}(\alpha_i-1) \Delta_{i}(t)\beta_i\alpha_i^{\Delta_i(t)} \nonumber\\& + (l_{i1}\alpha_i+l_{i2}\alpha_i - l_{i2})\beta_i\alpha_i^{\Delta_i(t)} \Big\},  \label{lyapunov-drift}
	\end{align}
	where $ \mathbb{E}[\gamma_i(t)|\Delta_i(t)] $ represents the conditional successful transmission rate under a scheduling policy based on AoI.
	
	Kim \emph{et al.}\cite{kim2014scheduling} proved that the Lyapunov drift (\ref{lyapunov-drift}) is minimized by greedily scheduling $ M $ sensors that have the largest $ G_i $s, i.e., the first term in the summation of (\ref{lyapunov-drift}) with $ \mathbb{E}[\gamma_i(t)|\Delta_i(t)] $ replaced by $ p_i $: 
	\begin{align}
		G_i = &{}p_i\beta_i\Big[l_{i1}\alpha_i\Delta_i(t)\alpha_i^{\Delta_i(t)} + (l_{i1} + l_{i2}) 
		\alpha_i(\alpha_i^{\Delta_i(t)} - 1)\Big].  \nonumber
		%	\label{equation-max-weight}
	\end{align}
	Furthermore, since $\forall i\in \mathcal{N} $, $ l_{i1}$ and $ l_{i2} $ are designable, by choosing them as follows: 
	\begin{subequations}
		\begin{align}
			l_{i1} &\triangleq \frac{p_i}{1-(1-p_i)\alpha_i},\label{app-l1}\\
			l_{i2} &\triangleq \frac{\alpha_i-1+p_i-2\alpha_ip_i}{(\alpha_i-1)[1-\alpha_i(1-p_i)]}, \label{app-l2}
		\end{align}\label{l}%
	\end{subequations}
	one can confirm that $ G_i $ becomes the same as the Whittle index in (\ref{W}). In this case, the policy that minimizes the Lyapunov drift (\ref{lyapunov-drift}) performs in the same way as our lightweight scheduling policy, and we can obtain the upper bound of $ J $ by analyzing the Lyapunov drift (\ref{lyapunov-drift}). 
	
	%	Given that the optimal policy that minimizes (\ref{lyapunov-drift}) is the greedy policy proposed by Kim \emph{et al.}\cite{kim2014scheduling}, the Lyapunov drift obtained based on the following randomized policy should be an upper bound of (\ref{lyapunov-drift}). 
	%	The randomized policy in this paper schedules the sensors as follows. Considering the constraint (\ref{optimal-problem-b}), there are totally $ \mathbb{C}^M_N $ choices of scheduling decisions, denoted by $ \{{\mathbf u}_j\in \mathcal{U}, j\in\{1,..., \mathbb{C}^M_N\}\} $. Thus, we schedule the sensors by randomly picking one of the choices with probability $ q^c_j=\mathbb{P}({\mathbf u}_j),  \forall j\in \{1, \cdots, \mathbb{C}^M_N\} $, where $ \sum q^c_j=1 $. This results in that the randomized policy schedules each sensor with a certain probability calculated as follows:
	%	\begin{align} 
		%		q_i=\sum_{j:u_{j,i}=1}q^c_j, \quad \forall i\in \mathcal{N}, \label{randomized-policy-q}
		%	\end{align}
	%	where $ u_{j,i} $ is the $ i $th element of $ {\mathbf u}_j $. Inversely, given $ \{q_i\} $, we can certainly find a feasible solution of the randomized policy $ \{p^c_j\} $ by solving (\ref{randomized-policy-q}) since the number of variables is less than that of equations.
	
	With (\ref{problem-statical-scheduler-c}) and (\ref{problem-statical-scheduler-b}), we have (\ref{upper-bound-existing-condition}). 
	Rearranging the terms in (\ref{inequality-lyapunov-drift}), we obtain 
	\begin{align} 
		\vec{\mathcal{L}}(t) + \sum_{i = 1}^N (\eta_i \Delta_i - \mathcal{S}_i) \beta_i\alpha_i^{\Delta_i(t)} \le 
		{} \sum_{i = 1}^N p_iq^{*}_i\beta_i\alpha_i(l_{i1} + l_{i2}). \nonumber 
	\end{align} 
	Further, considering (\ref{theorem-upperbound-h}) and (\ref{theorem-upperbound-c}), if $ \tilde{\Delta}_i^* > 1 $, then it is obvious that for $ \Delta_i\in [1, \tilde{\Delta}_i^*-1] $, 
	\begin{align}
		&\sum_{i=1}^N (\eta_i \tilde{\Delta}^*_i-\mathcal{S}_i)\beta_i\alpha_i^{\Delta_i} - \sum_{i = 1}^N (\eta_i \Delta_i-\mathcal{S}_i) \beta_i\alpha_i^{\Delta_i(t)} \nonumber\\
		\le{} &\sum_{i = 1}^N \eta_i (\tilde{\Delta}_i^* - \Delta_i)\beta_i\alpha_i^{\Delta_i} \nonumber\\
		\le{} &\sum_{i = 1}^N \eta_i \tilde{\Delta}_i^*\beta_i\alpha_i^{\tilde{\Delta}_i^*} = \mathcal{C},  \nonumber 
	\end{align} 
	which indicates that 
	\begin{align}
		\sum_{i=1}^N (\eta_i \tilde{\Delta}^*_i-\mathcal{S}_i)\beta_i\alpha_i^{\Delta_i} - \mathcal{C} \le \sum_{i = 1}^N (\eta_i \Delta_i-\mathcal{S}_i) \beta_i\alpha_i^{\Delta_i(t)}.  \nonumber 
	\end{align}
	Considering that $ \eta_i > 0 $, for $ \Delta_i\in [\tilde{\Delta}_i^*, \infty] $, the following inequality holds: 
	\begin{align}
		\sum_{i=1}^N (\eta_i \tilde{\Delta}^*_i-\mathcal{S}_i)\beta_i\alpha_i^{\Delta_i}\le{} \sum_{i=1}^N (\eta_i \Delta_i-\mathcal{S}_i)\beta_i\alpha_i^{\Delta_i}. \nonumber
	\end{align} 
	Thus, we can obtain
	\begin{align} 
		&\vec{\mathcal{L}}(t) + \sum_{i = 1}^N (\eta_i \tilde{\Delta}^*_i-\mathcal{S}_i) \beta_i\alpha_i^{\Delta_i(t)} - \mathcal{C} \nonumber\\ \le {} &\sum_{i = 1}^Np_iq^{*}_i\beta_i\alpha_i(l_{i1} + l_{i2}). \nonumber 
	\end{align} 
	Taking expectations on both sides of the above inequality and rearranging the terms generates: 
	\begin{align}
		\sum_{i = 1}^N (\eta_i \tilde{\Delta}^*_i-\mathcal{S}_i) \mathbb{E}\left[\beta_i\alpha_i^{\Delta_i(t)}\right] \le {} \mathcal{C} + \sum_{i = 1}^N p_iq^{*}_i\beta_i\alpha_i(l_{i1} + l_{i2}).  \nonumber
	\end{align} 
	Therefore, we obtain the upper bound (\ref{upper-bound}) by taking the minimum value of $\{ \eta_i \tilde{\Delta}^*_i-\mathcal{S}_i, i\in \mathcal{N} \} $. The proof is complete. 
	
	\subsection{Proof of Theorem \ref{theorem-necessary-condition-of-system-stability}\label{appendix-theorem-necessary-condition-of-system-stability}} 
	\begin{IEEEproof} 
		Given that $ J_{origin} < \infty $, we can conclude that each system $ i $ is stable. Additionally, we observe that the transmission decision $ \{u_i(t)=1, \forall t\in \mathbb{N}\} $ is the most stable for system $i$ under any scheduling policy (including our lightweight one). In other words, if system $i$ is not stable under the decision $ \{u_i(t)=1, \forall t\in \mathbb{N}\} $, it will not be stable under any other policies either. Based on this observation, we derive the necessary stability condition for system $i$ as given in (\ref{stability-condition-1}) \cite[Theorem 3]{you2011mean}, completing the proof. 
	\end{IEEEproof} 
	
	\subsection{Proof of Theorem \ref{theorem-sufficient-condition-of-system-stability}\label{appendix-sufficient-condition-of-system-stability}} 
	\begin{IEEEproof} 
		Since the randomized policy described in Appendix \ref{appendix-lyapunov-drift-inequality} renders an upper bound on the estimation error, a sufficient condition for the estimation stability under the optimal randomized policy is also sufficient for the stability under the proposed policy. 
		Under the randomized policy, we have
		\begin{align}
			\mathbb{P}(u_i(t))  = \left\{ 
			\begin{array}{ll}
				q_i^{*}, & \text{if }u_i(t) = 1, \\
				1-q_i^{*}, &\text{otherwise}. 
			\end{array}\right. \label{theorem-sufficient-condition-u}
		\end{align} 
		Substituting (\ref{theorem-sufficient-condition-u}) and $ \mathbb{P}(s_i(t) = 1) = p_i $ into (\ref{gamma-definition}), we derive the distribution of $ \gamma_i $ under the randomized policy as follows: 
		\begin{align}
			\mathbb{P}(\gamma_i(t))\left\{ 
			\begin{array}{ll}
				q_i^{*}p_i, &\text{if }\gamma_i(t) = 1, \\
				1-q_i^{*}p_i, &\text{otherwise}. 
			\end{array}\right. \label{theorem-sufficient-condition-gamma} 
		\end{align} 
		Considering (\ref{AoI}) and (\ref{theorem-sufficient-condition-gamma}), we further conclude that under the randomized policy, $ \Delta_i $ follows a geometric distributed, i.e., 
		\begin{align}
			\mathbb{P}(\Delta_i(t) = k) = q_i^{*}p_i(1-q_i^{*}p_i)^{k-1}. 
		\end{align}
		Thus, (\ref{stability-condition-2}) is a sufficient stability condition for system $ i $ under the randomized policy\cite[Theorem 12]{you2011mean}. 
	\end{IEEEproof}

\end{document}